# SPECTRE: A Hybrid System for an Adaptative and Optimised Cyber Threats Detection, Response and Investigation in Volatile Memory

Arslan Tariq Syed, Mohamed Chahine Ghanem *, Elhadj Benkhelifa, Fauzia Idrees Abro

***Abstract*—The increasing sophistication of modern cyber threats, particularly file-less malware relying on "living off the land" techniques, poses significant challenges to traditional detection mechanisms. Memory forensics has emerged as a critical approach to detecting such threats by analysing dynamic changes in system memory. This research introduces SPECTRE (Snapshot Processing, Emulation, Comparison, and Threat Reporting Engine), a modular Cyber incident response system designed to enhance threat detection, investigation, and visualization. By adopting Volatility's JSON format as an intermediate output, SPECTRE ensures compatibility with widely used Digital Forensics and Response (DFIR) tools, minimizing manual data transformations and enabling seamless integration into established workflows. Its emulation capabilities safely replicate realistic attack scenarios, such as credential dumping and malicious process injections, for controlled experimentation and validation. The anomaly detection module addresses critical attack vectors, including RunDLL32 abuse and malicious IP detection, while the IP forensics module enhances threat intelligence by integrating tools like Virus Total and geolocation APIs. SPECTRE's advanced visualization techniques transform raw memory data into actionable insights, aiding Red, Blue, and Purple teams in refining their strategies and responding more effectively to emerging threats. Bridging gaps between memory and network forensics, SPECTRE offers a scalable, robust platform for advancing threat detection, team training, and forensic research in combating sophisticated cyber threats.**

## I. Introduction

The landscape of cybercrime is advancing at an alarming pace, with adversaries increasingly leveraging "living off the land" and fileless techniques. These approaches exploit legitimate system tools, leaving minimal traces and significantly challenging traditional detection methods. As a result, memory acquisition and analysis have become essential components in investigating and mitigating these sophisticated attack strategies [1]. Fileless malware represents an escalating threat that poses significant challenges to traditional antivirus solutions [2]. Operating entirely in memory, it leaves no artifacts on the file system, enabling it to achieve a considerably higher success rate in evading detection compared to conventional malware types [3]. The comparison of memory snapshots plays a critical role in identifying malware by detecting changes such as newly created or modified processes, altered memory regions, or hidden code. These changes often indicate malware's efforts to persist or evade detection by manipulating memory. Through effective snapshot comparison, discrepancies can be uncovered, exposing suspicious activities and enhancing the detection of otherwise elusive malware [4].

### A. Research Scope

Based on the observations, the development of a novel system would fill a critical gap in existing security measures. Centred on efficient memory snapshot comparison, this system would strengthen the detection and analysis of sophisticated threats, significantly enhancing the capabilities of memory forensics in addressing contemporary cybercrimes. Such an approach aligns with the increasing demand for advanced detection techniques to counter the complexities of evolving attack methodologies. This research aims to advance memory forensics by designing a system capable of detecting and analysing malicious activities. The proposed system will enable the identification of potential threats by examining differences between memory snapshots and conducting in-depth analyses. By focusing on these objectives, the study seeks to provide a robust solution for uncovering sophisticated cyber threats that may evade traditional detection methods.

### B. Research Motivation and Questions

Modern cyber threats are increasingly exploiting system memory to bypass conventional file-based detection techniques. By capturing and examining memory snapshots, security analysts can uncover advanced malware concealed within memory. This paper focuses on enhancing the detection of such threats by integrating memory forensics, IP analysis, emulated test data, and advanced visualization [5]. The proposed system seeks to address gaps in existing solutions, enabling effective memory analysis, improved anomaly detection, threat scenario simulation, and robust protection. This research aims at bridging the gap in current detection and analysis of complex cyber threats through the introduction of a hybrid system. By leveraging memory snapshot analysis, visualization

* Dr Mohamed Chahine Ghanem the corresponding author
email: m.ghanem@londonmet.ac.uk.

Mr Arslan Tariq Syed is with Cyber Security Research Group. University of Westminster, London, UK.

Dr M.C. Ghanem is with the Cyber Security Research Centre. London Metropolitan University, London, UK. and Cybersecurity Institute, University of Liverpool, Liverpool, UK.

Prof. E. Benkhelifa is with the Smart Systems, AI and Cybersecurity Research Centre, Staffordshire University, Stafford, UK.

Dr F. Idrees Abro is with the Department of Information Security, Royal Holloway, University of London, Egham, UK.

techniques, and emulation of test data, the system is to provide a comprehensive solution for threat investigation. Referred to as SPECTRE, the system aim to answer the following research questions:

- **RQ1:** What open-source systems effectively represent memory snapshots interactively and support data integration?
- **RQ2:** How can memory snapshot differences reveal threats using indicators like processes, ports, and registries, and how can threat intelligence enhance detection?
- **RQ3:** What methods enable memory snapshot emulation for safe malware testing, dynamic modifications, and integration with forensic systems?
- **RQ4:** How can SPECTRE visualize memory snapshot deltas and support threat analysis through advanced visualization and emulation techniques?

### C. Article Organisation

The remaining parts of the paper are structured as follows; Section II provide a comprehesive Literature Review exploring existing methodologies, tools, and frameworks in memory forensics, highlighting research gaps. Section III Covers SPECTRE system Design and Implementation and provides details on SPECTRE's architecture, core components, plugin selection rationale, and implementation methods. It emphasises on Visualization and Anomaly Detection Modules showcasing their role in threat detection and analysis. Section IV Covers SPECTRE Results and Evaluation by assessing SPECTRE's performance in terms of scalability, efficiency and accuracy and Benchmarking with related works and industry solutions. V provide a comprehensive discussion of the obtained results and reflects on strengths, contributions, and limitations, while the Section VI provides a conclusion and summarizes the findings and suggests future improvements.

## II. Literature Review

The Literature Review section of this paper provides a comprehensive examination of existing research and methodologies in the fields of malware analysis, memory forensics, and detection techniques. This review highlights the evolution of key concepts, tools, and approaches used in detecting and mitigating sophisticated malware threats. The focus is on memory forensics, which has gained significant importance due to the growing complexity of modern cyber threats, particularly fileless malware that operates in volatile memory. The review also covers various analytical tools and frameworks, such as Volatility, which have played a crucial role in advancing memory analysis techniques. By synthesizing these findings, the review identifies key challenges and gaps in current methodologies, offering insights into future directions for improving malware detection and forensic investigations [6].

### A. Malware Analysis

Malware encompasses various forms of malicious software, including viruses, spyware, trojans, worms, and rootkits, designed to compromise computer systems by disrupting operations, exfiltrating data, or enabling unauthorized access. A critical component of malware analysis is the identification of suspicious behaviours, which plays a pivotal role in detecting, classifying, and clustering malware effectively [7]. Although behavior-based detection is crucial, relying exclusively on this approach may fail to identify novel malware that uses advanced evasion techniques. This underscores the necessity for incorporating complementary analysis methods to enhance detection capabilities. [8]

As technology continues to evolve in our increasingly digital world, the prevalence of malware has risen, highlighting the need for more effective protective measures. Traditional static and dynamic analysis methods often fall short in detecting sophisticated, modern threats. As a result, memory analysis-based approaches have gained greater importance in the ongoing battle against malware [9]. Despite the growing importance of memory analysis, it still faces challenges in identifying advanced threats that exploit memory in subtle ways. This emphasizes the need to combine memory analysis with other detection techniques to achieve more comprehensive threat detection.

Fileless malware is difficult to detect as it operates directly in RAM, bypassing traditional detection methods. It typically infiltrates systems through RAM access, phishing emails, or malicious websites, injecting itself into legitimate processes. While it often does not persist after a system reboot, attackers may employ techniques to maintain access [10]. Fileless malware is challenging to detect because it executes directly in RAM. However, its lack of persistence after a reboot presents a potential opportunity for mitigation.

VirusTotal is a widely-used platform that aggregates results from multiple antivirus engines, providing researchers with detailed malware analysis reports, labels, and threat classifications [11]. Integrating VirusTotal with other tools and techniques enhances the depth and thoroughness of malware analysis.

Stealthy malware often employs non-executable files, like .txt or .log files, which appear harmless but may contain encoded data that makes them dangerous [12]. Malware often hides within image files with standard extensions, like PNG or JPEG, making them appear benign while posing significant risks [13]. Memory forensics enables the detection of attacks by analysing running processes and identifying inconsistencies between file extensions and code behaviour, revealing concealed threats.

### B. Live Memory Forensics

Memory forensics has evolved from focusing solely on storage device analysis to becoming a key aspect of incident response, enabling the identification of malicious activities through the examination of volatile memory [14]. Memory forensics is essential in incident response; however, its integration into standard workflows remains difficult due to the complexity of volatile data and the requirement for specialized expertise.

Successful volatile memory analysis depends on accurate, interference-free memory dumps with integrity, but challenges like page smearing can undermine the quality of these captures [15]. Page smearing undermines data integrity, which in turn impacts the effectiveness of forensic analysis tools.

Memory analysis offers deep insights into processes, DLLs, and network connections, making it invaluable for identifying malware techniques like process and DLL hooking. It also reveals the system state and operating system details, surpassing traditional behavioral analysis by effectively monitoring rootkits and hidden threats. Its growing adoption underscores its effectiveness in detecting sophisticated malware [16]. Memory analysis is highly effective for malware detection but requires expert handling and advanced tools for optimal use. Memory analysis uncovers crucial evidence not found on hard drives, including network connections, IP addresses, timestamps, and malicious programs, making it essential for advancing digital forensics [17]. Memory analysis reveals transient forensic evidence critical for detecting malicious activities.

Attackers often exploit processes like cmd.exe, powershell.exe, wscript.exe, cscript.exe, and RunDLL32.exe in Windows OS for execution, persistence, and evasion tactics. Monitoring their parent-child relationships and analyzing command-line arguments can uncover signs of malicious activity, including revealing stolen credentials, IP addresses, and attack scripts. Network connections also provide critical information to identify potential attacker tools [1]. The emphasis on process monitoring and command-line analysis highlights key strategies for identifying malicious activity, providing a valuable basis for improving forensic methodologies

### *C. Memory Forensic Tools and Techniques*

Several memory analysis tools have been developed to assist users in extracting valuable artifacts from memory dumps. Open-source options include Volatility and Rekall, while commercial tools include Cellebrite Inspector, FireEye Redline, Magnet AXIOM, and WindowsSCOPE [15].

*1) Autopsy:* Autopsy is a forensic tool designed for analysing timelines and extracting web artifacts from data to reveal file history. It features data carving capabilities, enabling the recovery of deleted files that may still be present in memory. Additionally, by leveraging multiple cores, Autopsy can rapidly scan devices for signs of compromise through parallel processing [18].

*2) Redline:* Redline is an endpoint analysis tool designed for live memory response, focusing on enhancing endpoint memory security to prevent the exploitation of vulnerabilities. It is capable of detecting indicators of compromise and employs SHA-1 and MD5 encryption algorithms to ensure data integrity [18].

*3) Rekall:* Rekall, once a popular memory analysis tool, is now less relevant for future forensic research due to the cessation of its development [15].

*4) Volatility:* Volatility analyses memory dump files to extract detailed information. Its framework is organized by features such as processes, sockets, and handles, which are further categorized into more specific elements, like counting TCP and UDP sockets [18].

### *D. Volatility Framework*

The Volatility Framework is an open-source Python tool designed for analyzing RAM across multiple systems, including Windows, Linux, Mac, and Android. Its system-independent performance allows it to support new operating systems as they are introduced [19]. Volatility organizes extracted memory features into modules, enabling in-depth comparison and analysis [18]. This makes Volatility highly effective for extracting details from memory dumps, making it well-suited for the forensic analysis of malware.

Volatility can process various memory dump formats, including raw dumps, crash dumps, VMware (.vmem), and VirtualBox dumps. It also supports evidence verification by comparing hash values of stored, deleted, encrypted, or password-protected files using tools such as HashCalc [20]. Volatility efficiently analyses RAM and supports various dump formats; however, evidence verification requires the use of additional tools.

Volatility is widely used in both research and commercial sectors, with commercial tools offering added features that refine, rather than revolutionize, memory analysis methods [15]. The Volatility Framework extracts comprehensive data from memory dumps, including running processes, timestamps, and recoverable files. It can identify recently modified files, memory addresses, user browser history, active network sockets, and kernel modules. Additionally, it reveals file names of active processes, network connections, loaded libraries for each process, and virtual-to-physical address mappings [21]. While Volatility retrieves a wealth of data from memory dumps, its analysis and processing present significant challenges.

### *E. Data Formats and Standards*

JSON's simplicity, compatibility, and efficiency in data transfer have made it a vital tool for scalable data exchange, especially in APIs for major platforms [22]. VolMemLyzer is a Python script designed to extract features from memory dumps produced by Volatility. The output from each Volatility module is formatted as a table in JSON, which VolMemLyzer then writes to a temporary file [18], [23]. VolGPT, a Volatility-based tool, selectively leverages Volatility's JSON keys to tailor its functionality to specific needs [24]. Leveraging Volatility's JSON output can enable scalable and compatible software development, promoting seamless integration and data exchange across platforms.

### *F. Visualization*

Visualization methods are vital in malware detection, offering efficient ways to monitor and analyse malware behaviour using various graphical techniques like charts, graphs, scatter plots, and colour maps to identify unusual activity and attack patterns [25]. Visualization offers several key advantages, including faster detection of suspicious activities, clearer interpretation of complex data, enhanced decision-making, and the ability to monitor ongoing threats in real-time [26] and [27].

Data visualization simplifies complex information by converting it into graphical formats. Tools like spreadsheets provide basic charting, while advanced software such as Tableau offers interactive features [28]. Programming libraries like Matplotlib enable the creation of custom visualizations, though they require technical expertise. Selecting the appropriate tool and technique, whether bar charts, heat maps, or others is crucial for effective communication, depending on the data type and the intended audience [29]. Visualizing Volatility output can significantly enhance forensic analysis by converting raw data into intuitive graphical representations. This approach will allow analysts to quickly identify patterns and anomalies, thereby improving the accuracy and efficiency of investigations.

### G. Cyber Incident Data Emulation

Synthetic data mimics real data to ensure privacy and enable analysis, offering a valuable alternative in situations where real data is limited or privacy concerns arise [30]. Synthetic data allows for secure testing and development in sensitive areas by mimicking the characteristics of real data, even in cases where actual data is scarce or inaccessible [8]. This approach is particularly useful for benchmarking, profiling, and algorithm testing in fields such as cybersecurity and forensics, where data privacy and availability are often concerns [31].

To evaluate data-intensive algorithms, large and realistic datasets are often needed, and the Python library Faker serves as a useful tool for generating structured data for testing purposes [32]. Faker is a vital tool for generating realistic data, enabling more accurate testing and development. For instance, the open-source Python project RISC utilizes Faker to create realistic datasets by applying stochastic sampling with realistic distributions for various types of data, such as personal information, vehicle details, and contract record [33]. The Faker library is a crucial tool for generating synthetic data in memory forensic scenarios, enabling the creation of realistic process hierarchies, network connections, and malicious file names. Its ability to generate diverse datasets, such as dates, times, hostnames, and IP addresses, greatly improves forensic testing and analysis, providing security professionals with more accurate and comprehensive data for threat detection and response.

### H. Research Gaps

We describe briefly the main research gaps in memory forensics identified during this research.
*1) Delta Extraction for Malicious Activity Detection:* Detecting ongoing malicious activity requires analysing changes between memory snapshots, which static snapshots alone may not capture.
*2) Intermediate Format for Memory Snapshots and Deltas::* The lack of standardized formats for memory snapshots and deltas limits interoperability and automation in memory forensic processes.
*3) Enhanced Visualization for Snapshots and Deltas:* Advanced visualization methods are needed to effectively track and represent memory changes over time, aiding in anomaly detection and forensic analysis.
*4) Simulation Techniques for Memory Forensic Scenarios:* Realistic simulation environments, tailored specifically for memory forensics, can improve the testing and detection of malicious patterns.
*5) Automated Resource Integration in Memory Forensics:* Integrating online threat intelligence tools into memory forensic workflows can enhance the accuracy of threat assessment and detection.

## III. SPECTRE system Design and Implementation

### A. Overview and Components

SPECTRE is an advanced memory forensics system designed to enhance the detection and analysis of sophisticated cyber threats. It operates by analysing memory snapshots to identify anomalies and changes in system processes, network connections, and other system artifacts that may indicate malicious activity. Through its delta analysis methodology, SPECTRE compares memory states over time, uncovering subtle alterations that could signal an attack. The system also incorporates visualization techniques to track and represent these changes, making it easier for security analysts to detect unusual behaviours. SPECTRE's integration with existing forensic systems and its use of emulation techniques enables the safe testing of malware scenarios, providing a controlled environment for effective threat detection and investigative training. By combining memory and network forensics, SPECTRE offers a comprehensive, flexible solution for modern cybersecurity needs. SPECTRE architecture is shown in Figure 1

SPECTRE data flow diagram is shown in Figure 2

### B. Implementation

*1) Volatility Framework Rational:* Volatility selection rationale is shown in Table I
*2) Alternative Memory Forensic systems:* Reasons for discarding alternative tools are shown in Table II

*3) Volatility Plugin vs Standalone Utility:* Volatility can be used for memory analysis either through tight integration with its plugin system, offering community support and efficient workflows, or by utilizing its outputs in a custom utility, which provides more control but requires more effort. For SPECTRE, which involves snapshot comparison and future flexibility, a separate utility is better suited for innovation and customization.

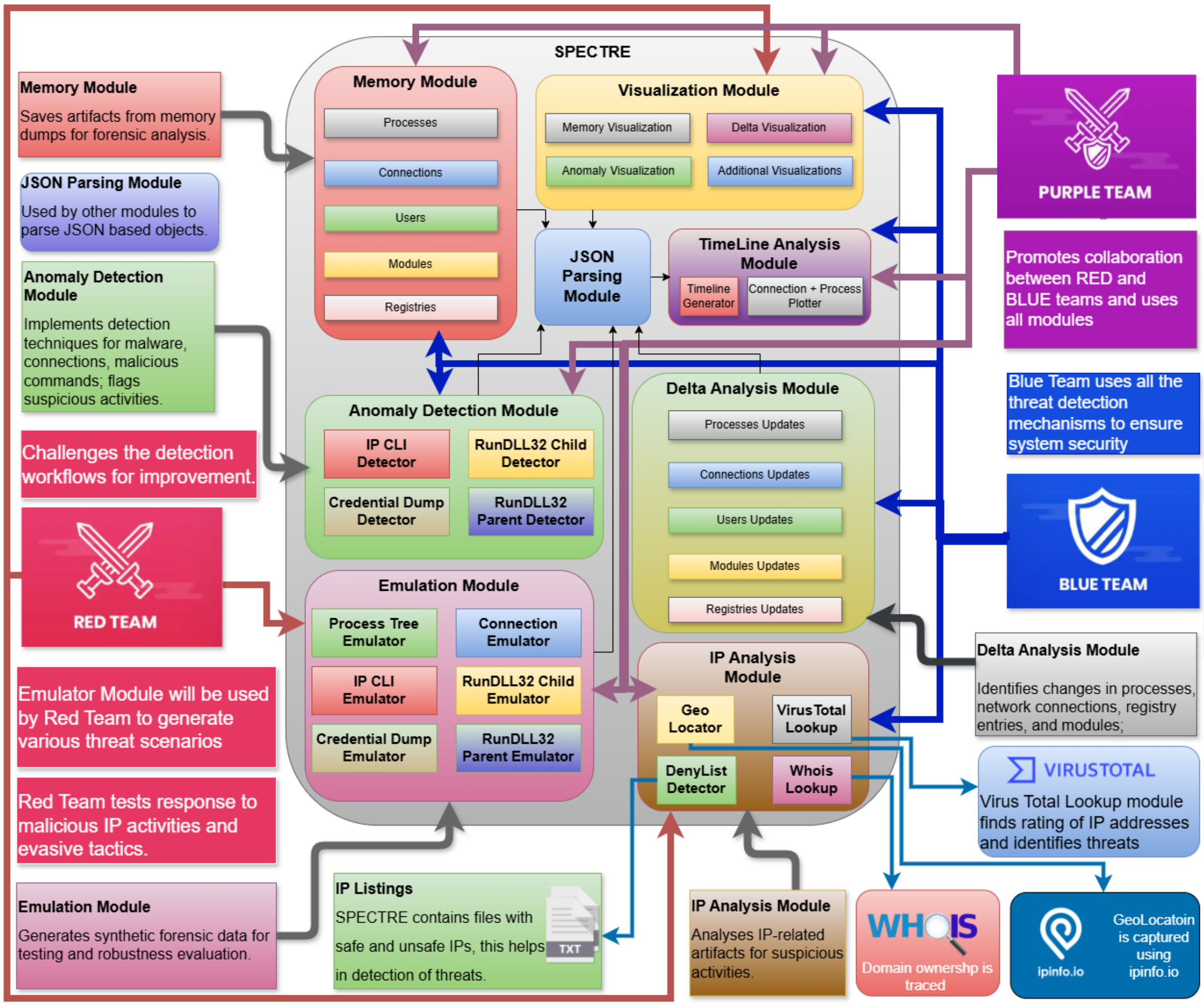

**Fig. 1:** SPECTRE System Overall Architectural Design

*4) Volatility JSON output as an Intermediate Format:* Volatility supports JSON as an output format [36], which is widely utilized across various systems. Some, like VolMemLyzer, fully leverage the JSON structure [23], while others, such as volGPT, use specific keys for their purposes [24]. JSON serves as an excellent intermediate format for memory forensic applications due to its versatility, supporting seamless integration and interoperability. It is also ideal for web interfaces, databases, and platforms like Elastic Search and Kibana [36]. The format follows the ISO/IEC 21778:2017 standard [37].

**Advantages:** JSON is a lightweight, structured, and language-agnostic format that facilitates efficient data transfer, sorting, and readability [22].

**Disadvantages:** JSON can be verbose, and for more concise data needs, alternative formats may be better [22]. However, its verbosity does not affect SPECTRE, since it leverages JSON's readability and schema-less design, ensuring compatibility and flexibility for memory forensics.

**Limitations:** JSON has limited support for comments, is vulnerable when used with untrusted services, and lacks a native date type [22]. These limitations can be addressed by using external documentation for comments, validating data to handle security concerns, and customizing date handling as needed [22].

*5) JSON Alternatives:* **Protocol Buffers:** A high-performance, language-agnostic format developed by Google for efficiently serializing structured data, reducing response times and payload size [38], [39]. Despite its advantages, it is not human-readable [40], requires external systems for interpretation [41], and has limited support in non-object-oriented languages [42]. Additional issues include complications with equality checks due to multiple binary formats for same data [42], lack of self-descriptiveness [42], and the need for special memory handling for large messages [43]. Moreover, it may not comply with legal standards [42]. **XML:** XML is a human-readable format used for flexible data structuring and custom syntax creation [22]. However, it is less efficient than JSON in terms of speed, memory usage, and power consumption [40], [44]. Protocol Buffers and XML provide alternatives to JSON but have significant drawbacks in terms of readability, efficiency, and suitability for memory forensic workflows.

## *C. Volatility Plugins in SPECTRE*

*1) Processes:* **PsTree Plugin**: Displays processes in a hierarchical structure [17], revealing parent-child relationships [45],

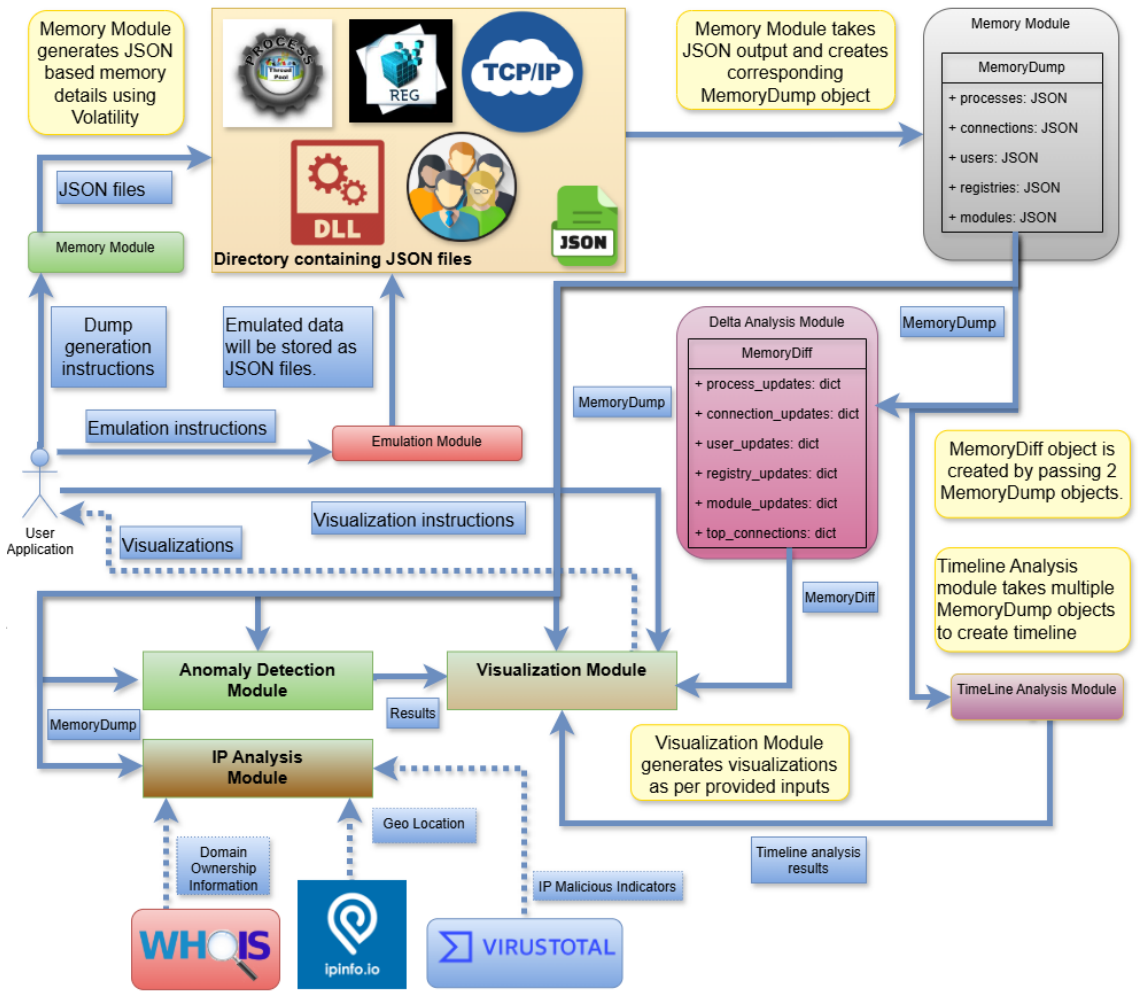

**Fig. 2:** SPECTRE Data Flow Diagram

**TABLE I:** Volatility Framework Selection.

| Aspect | Description |
|---|---|
| Rational | Open-Source Python tool, flexible, community-supported, and easily modifiable for diverse applications [19]. Command-Line Execution: Enables greater control, automation, and faster performance than graphical interfaces [7]. Cross-Platform Compatibility: Supports multiple operating systems and platforms [1]. Memory Dump Support: Compatible with multiple OS memory dumps [19] and various dump formats [20]. Modular Design: Features are organized into distinct modules, enabling granular and precise analysis [18]. Research and Commercial Integration: Widely used in research and integrated into commercial forensic tools due to flexibility, strong community support, and reliability [15]. |
| Advantages | Retrieves extensive information including processes, timestamp of the captured image, files, memory addresses, browser history, kernel modules, network connections [21], handles, TCP and UDP connections [18], user activities, application status, credentials, registries, symbolic links, command line arguments, abnormal behaviour, linked libraries, scheduled tasks, timelines [1]. |
| Disadvantages | No built-in support to compare differences in memory dumps. No built-in support to visualize information over the timeline for multiple snapshots. |
| Limitations & Mitigations | No built-in graphical user interface support [15]. Utilized Volatility Workbench, a graphical user interface (GUI) for Volatility [34], to enhance understanding of the system [1]. The Volatility Workbench is regularly updated to support the latest Volatility versions, including Volatility 3. However, it lacks full support for all Volatility parameters. |

**TABLE II:** Alternative Tools and Systems

| Tool | Description | Reason for disregarding |
|---|---|---|
| Autopsy | Autopsy utilizes Volatility plug-ins for analysing memory image files, effectively serving as a user-friendly graphical interface for Volatility memory forensics [7]. | Autopsy consumes nearly three times the percentage of memory compared to Volatility [7]. This makes it unsuitable for low end machines.<br>Underlying command line tool, The Sleuth Kit (TSK) is no longer being actively developed [7]. It limits enhancements of software scripts based on TSK. |
| Redline | Redline is a forensic toolkit designed for detecting malicious activity by analysing memory and files. It gathers comprehensive data, such as processes, drivers, and network information, to assess potential threats effectively [35] | Redline is the least developed tool in this area [35], [7].<br>It displays handles for terminated malware processes but lacks details like PID, creation time, and exit time [7].<br>Compared to Volatility and Autopsy, Redline's ability to detect injected malicious processes is limited [7].<br>Additionally, Redline consumes twice the memory percentage of Volatility [7]. Its versatility falls short when compared to Volatility [7].<br>Furthermore, Redline's CPU resource consumption is significantly higher than that of Autopsy and Volatility [7]. |

useful for detecting suspicious processes,

**PsList Plugin**: Lists active processes with details like PID and start time [45], plugin output truncates process names.

**Reason for Selection**: **PsTree** is preferred due to its ability to show full process names along with additional details, making it more effective for identifying malicious processes. This is evident in Figures 4 and 3.

*2) Threads:* The **pstree** plugin output includes thread count, as shown in Figure 3, making it useful for detecting threads as well.

*3) Command Line Arguments:* The **pstree** plugin displays command line arguments under the Cmd key, as shown in Figure 5, allowing pstree to be used for detecting command line arguments as well.

```
{
    "Audit": "\\Device\\HarddiskVolume4\\Users\\user\\Downloads\\
    msys2-x86_64-20240727.img",
    "Cmd": "C:\\Users\\user\\Downloads\\msys2-x86_64-20240727.img ",
    "CreateTime": "2024-09-22T01:57:09+00:00",
    "ExitTime": null,
    "Handles": null,
    "ImageFileName": "msys2-x86_64-2",
    "Offset(V)": 243575898931328,
    "PID": 14712,
    "PPID": 2612,
    "Path": "C:\\Users\\user\\Downloads\\msys2-x86_64-20240727.img",
    "SessionId": 1,
    "Threads": 7,
    "Wow64": false,
    "__children": [
```

**Fig. 3:** pstree json output

```
{
    "CreateTime": "2024-09-22T01:57:09+00:00",
    "ExitTime": null,
    "File output": "Disabled",
    "Handles": null,
    "ImageFileName": "msys2-x86_64-2",
    "Offset(V)": 243575898931328,
    "PID": 14712,
    "PPID": 2612,
    "SessionId": 1,
    "Threads": 7,
    "Wow64": false,
    "__children": []
}
```

**Fig. 4:** pslist json output

```
"Audit": "\\Device\\HarddiskVolume4\\Windows\\System32\\
csrss.exe",
"Cmd": "%SystemRoot%\\system32\\csrss.exe ObjectDirectory=\\
Windows SharedSection=1024,20480,768 Windows=On
SubSystemType=Windows ServerDll=basesrv,1
ServerDll=winsrv:UserServerDllInitialization,3
ServerDll=sxssrv,4 ProfileControl=Off MaxRequestThreads=16",
"CreateTime": "2024-09-16T03:46:57+00:00",
"ExitTime": null,
```

**Fig. 5:** Command Line in pstree json output

*4) Users:* **Selected Volatility Plugin**: **Hashdump** retrieves local user account names, relative identifiers (RIDs), and both LM and NT hashes [1]. A sample output is shown in Figure 6.

**Alternative Plugin**: **Cachedump** provides another method for extracting credentials, retrieving hashes of cached passwords for domain users [1].

**Selection Reasoning**: Based on practical observation, **Hashdump** consistently generated the user list, whereas **Cachedump** failed to retrieve credentials in certain snapshots, resulting in empty responses. Therefore, **Hashdump** is used to capture user information.

```
{
    "User": "Administrator",
    "__children": [],
    "lmhash": "aad3b435b51404eeaad3b435b51404ee",
    "nthash": "31d6cfe0d16ae931b73c59d7e0c089c0",
    "rid": 500
},
{
    "User": "Guest",
    "__children": [],
    "lmhash": "aad3b435b51404eeaad3b435b51404ee",
    "nthash": "31d6cfe0d16ae931b73c59d7e0c089c0",
    "rid": 501
}
```

**Fig. 6:** hashdump json output

*5) Network Connections and Ports:* **Main Volatility Plugin**: The **netstat** plugin in Volatility 3 offers detailed network data, such as connection states, process associations, and full IPv4/IPv6 support by accessing kernel data structures [46].

**Alternative Plugin**: The **netscan** plugin analyzes network connections, providing information such as protocol, IP addresses, ports, PID, process name, creation time, and connection state. It is useful for detecting suspicious network activity and tracking processes involved [1].

**Selection Reasoning**: Combining outputs from both **netstat** and **netscan** plugins is crucial for a complete network connectivity analysis. **Netstat** excels in network analysis by directly accessing kernel data structures [46], capturing detailed connection states and associating connections with processes while supporting both IPv4 and IPv6 protocols. In contrast, **netscan** uses memory scanning [47], which, although effective, may miss some connections due to memory corruption or incomplete metadata. However, in one sample, it is observed that **netscan** detects port 0 connections, which **netstat** overlooks. Recent studies have shown that port 0 traffic, initially associated with OS fingerprinting, is now primarily linked to large-scale DoS attacks and network scanning, where scanners probe networks in a non-repetitive pattern [48], [49]. In a practical observation, **netscan** missed about 38% of connections, while **netstat** missed UDP connections (ports 0 on UDPv4 and UDPv6) that were detected by **netscan**. Thus, using both plugins together enhances the reliability and comprehensiveness of network assessments.

*6) DLLs:* **Selected Volatility Plugin**: The **LDRModules** plugin identifies DLLs associated with processes [1], this plugin detects suspicious DLLS by comparing the Process Environment Block (PEB) in user space with Virtual Address Descriptors (VADs) in kernel memory [50].

**Alternative Plugin**: The **DllList** plugin lists the libraries loaded within a process [1].

**Selection Reasoning**: The **LDRModules** plugin is preferred over **DllList** because it offers a more thorough analysis. It examines all three module management lists: LoadOrderList (tracks module load order), MemoryOrderList (tracks module mapping in memory), and InitOrderList (tracks execution order of DllMain). On the other hand, **DllList** relies only on the LoadOrderList, which can be manipulated by attackers to conceal libraries [1].

### *D. SPECTRE Requirements and Dependencies*

SPECTRE is developed using Python version 3 and was implemented on a Windows 10 environment. It currently supports analysis of RAM images specific to Windows-based systems. The additional Python libraries required to run SPECTRE, which are not included by default in Python 3.7.3, are outlined in the Table III.

### *E. SPECTRE Visualization*

Matplotlib was chosen for SPECTRE visualization due to its extensive features, customizability, cross-platform compatibility, and interactivity. Seaborn was excluded due to frequent memory issues and poor integration with Python. Plotly was considered unsuitable because of its lack of automation, dependency on multiple libraries, slower performance, and high resource use [51]. Excel was excluded for lacking interactivity, whereas Matplotlib offers real-time, interactive integration with SPECTRE.

**TABLE III:** SPECTRE Library Requirements

| Library | Description | Installation Command |
|---|---|---|
| matplotlib | A comprehensive library for creating static, animated, and interactive visualizations in Python. | `pip install matplotlib` |
| Faker | A library for generating fake data such as names, addresses, emails, and phone numbers. | `pip install faker` |
| requests | A library for sending HTTP requests to access web resources. | `pip install requests` |
| dns.resolver | A module from dnspython for querying DNS records. | `pip install dnspython` |
| networkx | A library for creating, analyzing, and visualizing complex networks or graphs. | `pip install networkx` |
| PrettyTable | A simple library for rendering ASCII tables. | `pip install PrettyTable` |
| colorama | A library for producing colored terminal text and styles. | `pip install colorama` |
| tabulate | A library for pretty-printing tabular data in plain text, Markdown, and other formats. | `pip install tabulate` |
| Pillow | A fork of the Python Imaging Library (PIL) for image processing. | `pip install pillow` |
| seaborn | A statistical data visualization library based on Matplotlib. | `pip install seaborn` |
| scipy | A library for scientific computing and technical computing. | `pip install scipy` |

### F. Anomaly Detection Algorithms

*1) RunDLL32.exe based detections:* RunDLL32.exe is a trusted Windows executable that loads DLLs, enabling it to bypass security controls like AppLocker and Software Restriction Policies [52]. It is also used by attackers to dump process memory, including LSASS, for credential theft [52].

**Malicious RunDLL32 Child Processes**

Detecting malicious behavior requires understanding normal activity. Monitoring RunDLL32 executions, especially those triggered by unusual parent processes, is key. The algorithm [53] in Figure 7 can help identify abnormal RunDLL32 activity, aiding in early threat detection.

**RunDLL32.exe Suspicious Behaviour Detection**

This algorithm in Figure 8, flags RunDLL32.exe when it runs without command-line arguments, especially if it spawns child processes or opens network connections. Since RunDLL32.exe typically operates with specific parameters, these activities are likely indicators of malicious use [53].

*2) Credential Dumping:* Threat actors often exploit credentials stored in Windows LSASS to perform lateral movement and privilege escalation within networks, gaining extensive control over systems [54]. ProcDump, a command-line tool for monitoring and diagnostics [55], is frequently abused in credential theft attacks [54]. For instance, attackers can dump LSASS memory using command 'procdump -ma lsass.exe lsass_dump' [56]. Additionally, tools like 'comsvcs.dll' with 'RunDLL32.exe' enable similar memory-dumping operations, highlighting how attackers misuse system processes to extract credentials [56]. The credential dumping detection algorithm is depicted in Figure 9.

*3) Malicious Extension:* Malicious extension detection algorithm is shown in Figure 10. This algorithms uses pstree output to detect the process extension.

### G. Emulation Methodology

The research uses test data emulation to mimic real process behaviours, aiding anomaly detection and providing a controlled environment for training and validating security responses.

*1) Techniques of JSON-Based Emulation:* SPECTRE generated JSON-formatted process and connection data, replicating Volatility outputs and supporting malware scenario simulations through anomaly generation techniques. The next section explores the emulation of processes and connections, followed by three techniques that demonstrate SPECTRE's flexibility in creating anomalies.

**Process Tree Emulation** To emulate the given `pstree` JSON data with SPECTRE, the following steps were employed:

- **Process Names**: Process names were generated using `faker.Faker`, resulting in realistic executable names (e.g., `antivirus-1234.exe`, `scanner-5678.exe`).
- **PID and PPID**: Root processes were assigned unique PIDs (e.g., `1001`) with a PPID of `0` or user-specified values. Child processes were given new PIDs (e.g., `1002`) linked to their parent's PID.
- **Timestamps**: Random `CreateTime` values were assigned to each process (e.g., `2024-10-20T08:30:00`).
- **Process Attributes**: Attributes such as `Handles`, `ImageFileName`, `Offset(V)`, `Path`, `SessionId`, `Threads`, and `Wow64` were set (e.g., `C:\Program Files\Windows\antivirus-1234.exe`, with `5 threads`).
- **Child Processes**: Child processes were created under the root, ensuring correct `PPID` linkage.
- **JSON Output**: The processes and their hierarchy were structured and outputted in JSON format.

**Connection Emulation**

To emulate the Volatility `netstat` output in SPECTRE, the following steps were employed:

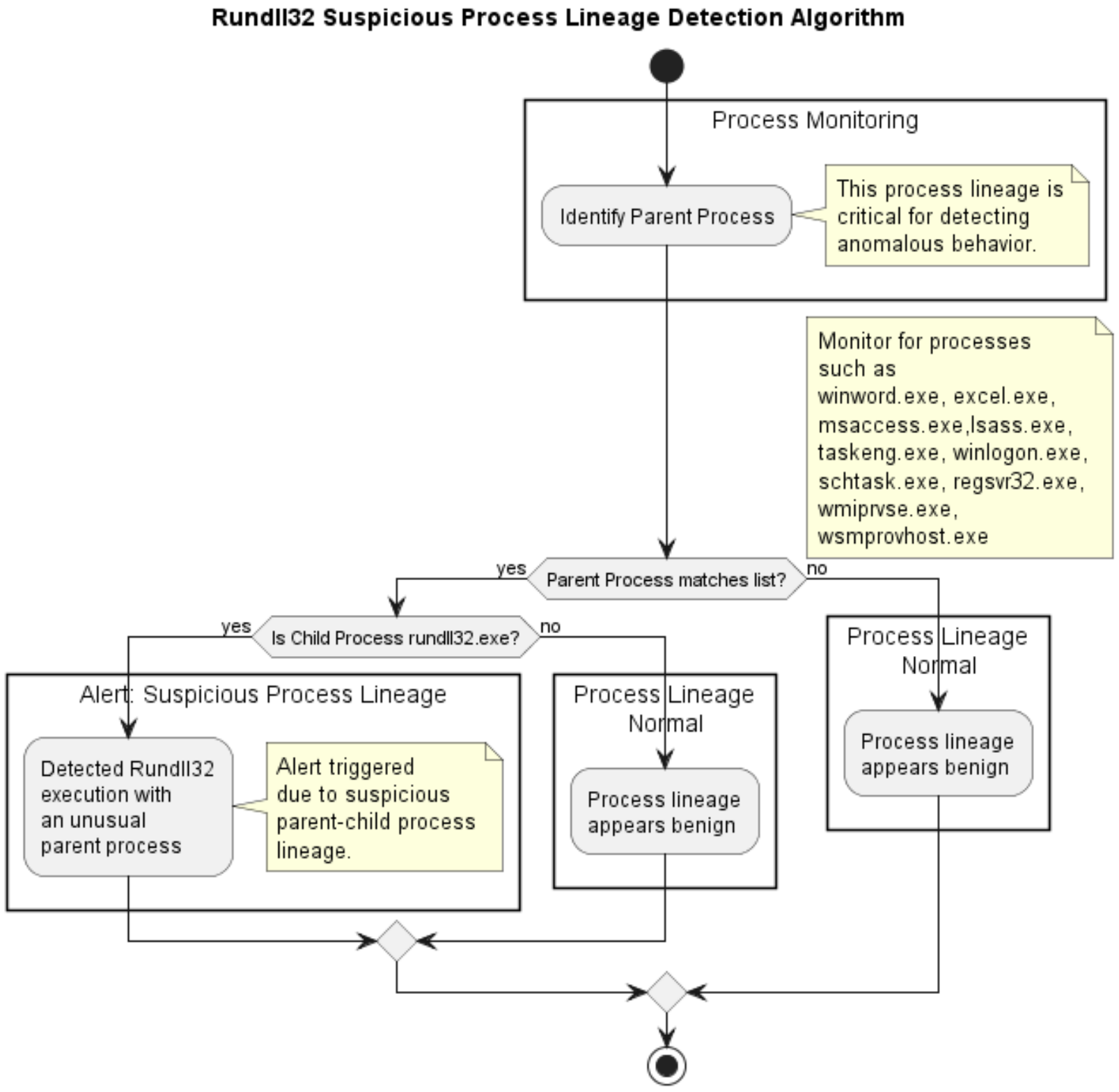

**Fig. 7:** Malicious RunDLL32.exe child process

- Foreign addresses (`ForeignAddr`) were chosen randomly from benign or malicious IP lists to simulate different scenarios.
- Ports (`ForeignPort` and `LocalPort`) were randomly generated using `random_int(min=1024, max=65535)`.
- Connection states were randomly selected from a list of known valid states.
- Process IDs (`PID`) were selected from the process tree, establishing ownership between processes and network connections.
- Process names were derived from the corresponding `pstree.json` data.
- Memory offsets were simulated using `random_int(min=1, max=999999999999999)`
- Protocols (`Proto`) were randomly selected from `['UDPv4', 'UDPv6', 'TCPv4', 'TCPv6']` to create diverse network scenarios.

This approach enabled SPECTRE to simulate processes listening on dynamically generated ports, connected to realistic IP addresses, with protocols and ownership linked to processes from `pstree.json`. SPECTRE provided command-line options to customize connection generation, allowing users to define the number of connections, specify benign and malicious IP lists, control their ratio, and link connections to valid processes. These features enabled flexible and realistic connection emulations.

*2) Malware Scenario Emulations:* Anomaly detection algorithms are inverted to develop the corresponding emulation algorithms. These emulation techniques use pstree and netstat emulators to generate malicious, benign and low risk scenarios. These scenarios are shown in Figure 11, Figure 12, Figure 13 and Figure 14.

### H. Network Forensics

*1) IP Geolocation:* We performed geolocation using ipinfo.io to identifies the geographic location, helping to link IPs with regions prone to cybercrime or suspicious activity. Pinfo delivers high accuracy, identifying city-level locations for 89% of targets, outperforming RIPE Atlas (73%) and MaxMind (55%) [57].

*2) Domain Lookup:* Purpose: Discovers network operators and responsible entities.
Method: WHOIS lookup.
Justification for Choice: WHOIS was selected over RDAP for its faster, more reliable performance. RDAP's deeper, recursive structures complicate parsing, and its larger size offers redundant data without additional value [58].

*3) Blacklist Check:* Purpose: Determines if the IP is associated with spam, malware, or DDoS attacks.
Method: VirusTotal Threat Intelligence.
Justification for Choice: VirusTotal provides comprehensive and actionable cyber threat intelligence on IP addresses [59]. During testing, it was found that VirusTotal accurately detects both malicious and benign IPs, while Hybrid-Analysis failed to provide relevant data for the tested IPs, confirming VirusTotal's suitability for the system.

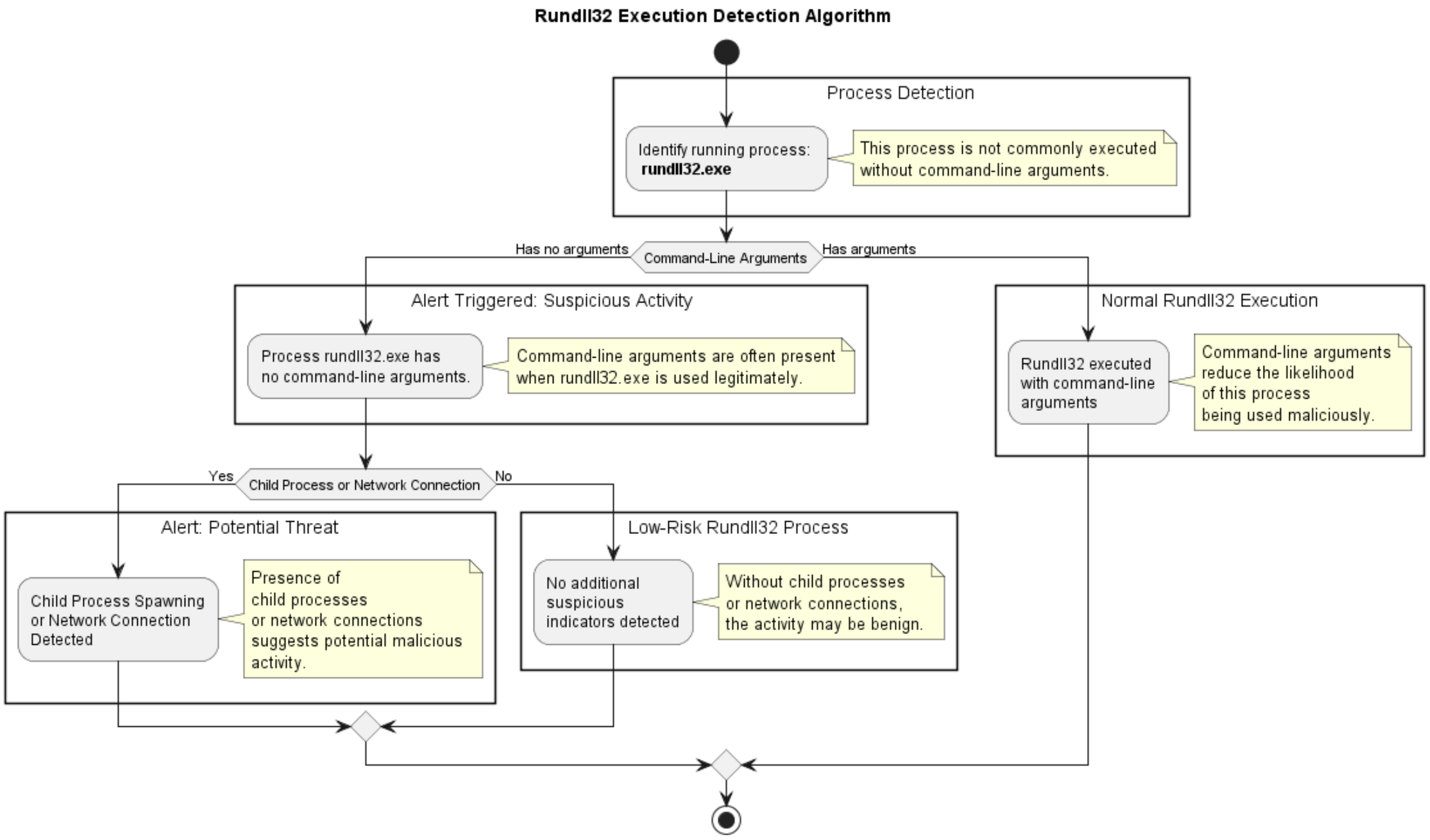

**Fig. 8:** Malicious RunDLL32.exe process

### *I. Analysis Methods*

*1) Memory Analysis Details:* The memory analysis captures key system data, such as processes, network connections, users, modules, IP analysis, and credential dumps, using volatility plugins like 'pstree', 'netstat', 'hashdump', and 'ldrmodules'. This data is analyzed to detect anomalies, unauthorized access, suspicious network activities, DLL injections, and credential dumping attempts.

*2) Delta Analysis:* Delta analysis monitors changes in system entities (e.g., processes, network connections, DLLs, registry keys, and users) over time. These changes—such as additions, removals, updates, and consistencies—are visualized to detect anomalies and potential malicious activities. These delta calculations provide insights into suspicious activity, such as the appearance of a potentially malicious connection from RunDLL32.exe.

*3) Timeline Analysis:* Timeline analysis tracks fluctuations in the state of system entities (e.g., processes, connections) over time, identifying deviations from normal trends for further scrutiny. For example, netstat output is analyzed over time to detect network behavior changes.

The process involves comparing multiple snapshots, categorizing connections into added, removed, updated, and consistent statuses, which are then visualized for further analysis. Changes in network behavior, such as new connections or state transitions, are tracked over time to detect attacks or malicious activities.

## IV. SPECTRE Deployment and Performances Evaluation

*1) Evaluation Metrics:* The following outlines the key metrics used to evaluate the SPECTRE system, highlighting their purpose and importance:

**Throughput:** Measures the volume of data processed by SPECTRE within a specific time frame. This metric demonstrates the system's ability to handle large datasets efficiently, particularly in modules such as emulation and delta analysis.

**Latency:** Assesses the time taken to process inputs and deliver outputs. Low latency is critical for time-sensitive operations, including memory analysis and timeline generation, ensuring rapid threat detection.

**Accuracy:** Evaluates the precision of detection algorithms in identifying anomalies and threats. High accuracy is essential to correctly identify threats while minimizing false positives, particularly in anomaly detection and delta analysis.

**Scalability:** Examines the system's performance with increasing input sizes. Scalability ensures that larger datasets can be processed efficiently without disproportionate resource usage, making it suitable for extensive emulation and memory analysis tasks.

**Resource Utilization:** Monitors CPU, memory, and storage usage to assess efficiency. Effective resource utilization is crucial to maintain system stability and optimize performance in resource-intensive modules like memory and anomaly analysis.

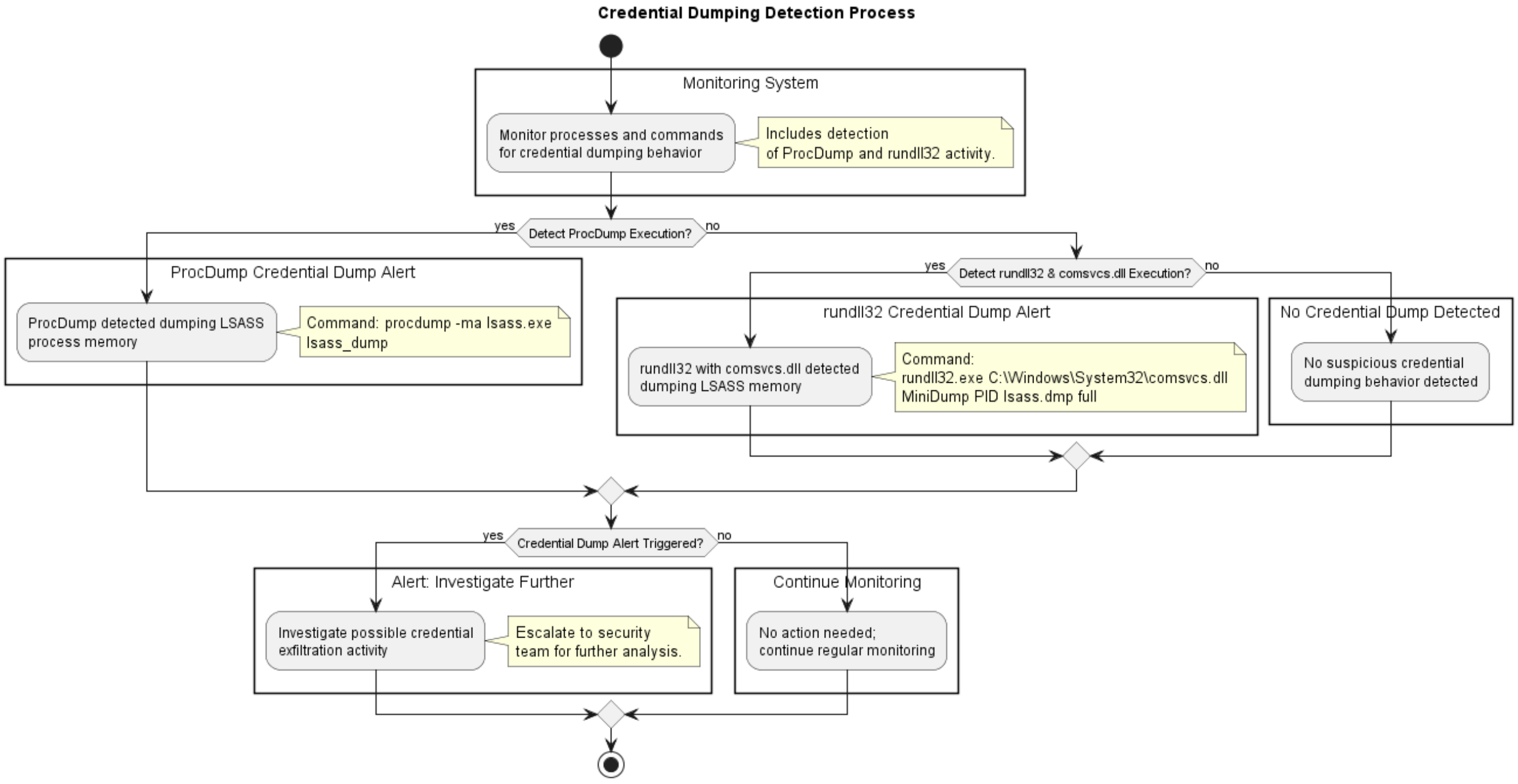

**Fig. 9:** Credential Dumping Detection

### A. *Data Collection*

This research employs three data collection modes tailored to forensic requirements:

- **Online Memory Dumps:** Real-world datasets from repositories test SPECTRE's capabilities in handling complex, realistic scenarios.
- **RAM Data via FTK Imager:** Controlled data from live systems validates accuracy and specific scenarios.
- **Automated Test Data:** Synthetic datasets generated with Python's Faker library stress-test edge cases and module robustness.

### B. *Non-Malware Test Data*

Real malware data was avoided in favour of synthetic datasets due to safety and legal concerns, availability of adequate alternatives, and the focused goals of the study. Emulated datasets provided realistic and relevant conditions, aligning with the system's purpose and enabling safe, customizable testing without the risks of real malware.

### C. *Automated Test Data*

The emulated test data incorporates the following combinations of memory snapshots:

*1) Automated Benchmarking Scenarios:*

- These scenarios involve $n$ values of 10, 100, 500, 5000, and 10,000, with the following attributes:
  - Between $n$ and $n \times 1.5$ processes.
  - $n \times 2$ connections associated with the processes.
  - $n$ modules linked to the processes.
  - $n$ users and $n$ registry entries.
- These configurations are designed to benchmark system performance in terms of time and memory utilization.
- The scenarios are employed to test key functionalities, including Memory analysis, Anomaly detection, Delta analysis, Combined plotting of connections, and processes and Timeline analysis.

*2) Other Emulation Modules:* Additional modules are utilized to simulate all supported malicious scenarios.

This comprehensive emulation framework supports rigorous testing across multiple threat and system performance dimensions.

### D. *Testing and Results*

*1) Memory Forensic System:* SPECTRE system enables efficient memory analysis with advanced visualization, threat detection, timeline analysis, and memory emulation for realistic scenarios. Figure 15 illustrates the integration of these components to facilitate efficient memory forensics and threat detection.

*2) SPECTRE Plots:* SPECTRE generates specialized plots that highlight key elements of memory images, focusing on critical aspects to enhance the analysis. Memory Analysis plot is shown in Figure 16, Figure 17 shows the Anomaly Analysis plot, Figure 18 demonstrates the Processes Scatter Plot, Figure 19 shows the Delta Analysis plot, and Figure 20 depicts Connections and Processes timeline plot.

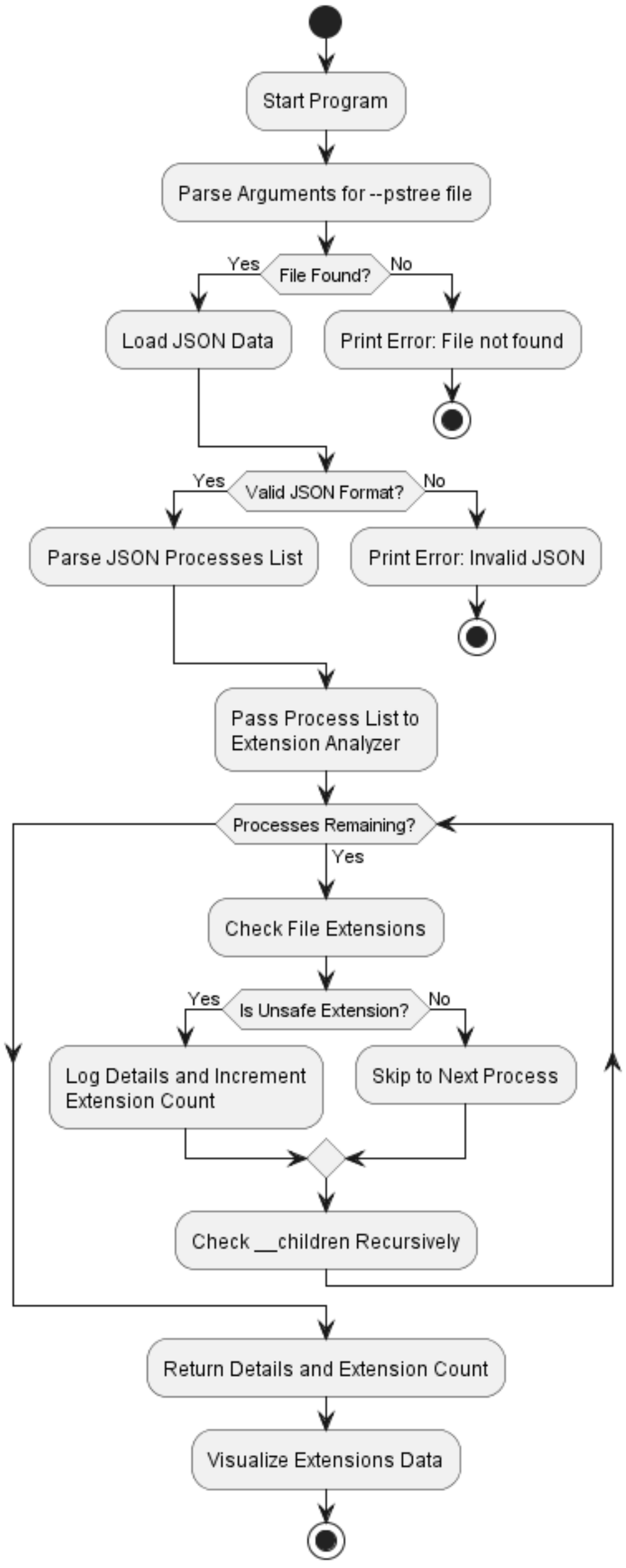

**Fig. 10:** Malicious Extension Detection

*3) Test Results:* **Emulation Classes Test Results:** The emulation classes demonstrated overall reliability and accuracy, successfully passing tests that validated their ability to generate realistic forensic scenarios and handle diverse functionalities, including process emulation, command generation, and data consistency. These results confirm their robustness for use in forensic analysis. In total, 35 tests were conducted across six modules, all of which were successfully passed.

**Anomaly Detection Classes Test Results:**

The anomaly detection modules exhibited strong functionality and seamless integration, successfully passing tests that validated their ability to identify malicious behaviours, analyze processes and connections, and leverage external intelligence for enhanced forensic insights. These results affirm their effectiveness in identifying and categorizing threats in diverse scenarios. Overall, 55 tests were conducted across seven modules, all of which passed successfully, covering both automated and real-world data types.

**Memory Analysis Tests Results:**

The system demonstrated comprehensive capabilities in memory analysis, delta analysis, and timeline plotting, successfully passing extensive tests that verified data handling, report generation, and visual representation. These results highlight its reliability and precision in processing and analyzing diverse memory and timeline scenarios. In total, 166 tests were executed across four modules, all passing successfully, with both automated and real-world data inputs were utilized.

*4) Benchmarking Results:* The benchmark results for time and memory are shown in Figure 21 and Figure 22, respectively.

*5) SPECTRE Emulation:* The use of Volatility's JSON format for memory snapshots enhances compatibility and streamlines integration with forensic tools like VolMemLyzer and volGPT. This standardized format reduces manual effort, boosts operational efficiency, and reinforces SPECTRE's utility in both simulated and real-world forensic investigations.

**Emulation Trade-offs** The JSON intermediate format enhances interoperability but introduces dependencies, potentially leading to compatibility issues if Volatility's output changes. Addressing this requires ensuring adaptability to evolving formats. Overall, SPECTRE's results highlight its transformative role in memory forensics, with attention to interoperability in dynamic contexts.

*6) SPECTRE Visualizations:* SPECTRE's visualizations enhance cybersecurity analysts' efficiency and accuracy by providing detailed insights into system behaviour through memory, delta, and timeline analysis plots. These tools support proactive threat detection and response.

*7) Visualizations and Analysis Plots:* SPECTRE's visualizations and plots offer comprehensive insights into memory and anomaly analysis. Memory analysis plots identify process and connection anomalies, such as irregular spikes, excessive child processes, and unusual protocol usage, while highlighting potential threats like DLL injections and thread hijacking. Anomaly detection plots visualize unsafe extensions, malicious process hierarchies, and foreign connection patterns, enabling early detection of high-risk entities and attack vectors.

Scatter plots provide dynamic, time-based views of process activity, with features like interactive tooltips and colour-coded classifications enhancing user engagement. These plots detect connection spikes and anomalous behaviours over time. Delta analysis plots track changes across memory snapshots, uncovering anomalies like zero-day attacks and correlating updates to advanced threats. Processes and connections timeline plots map connections and activities over time, identifying temporal anomalies such as off-peak DDoS activity and tracing the origins of persistent threats. Together, these visualizations ensure effective, time-sensitive forensic analysis.

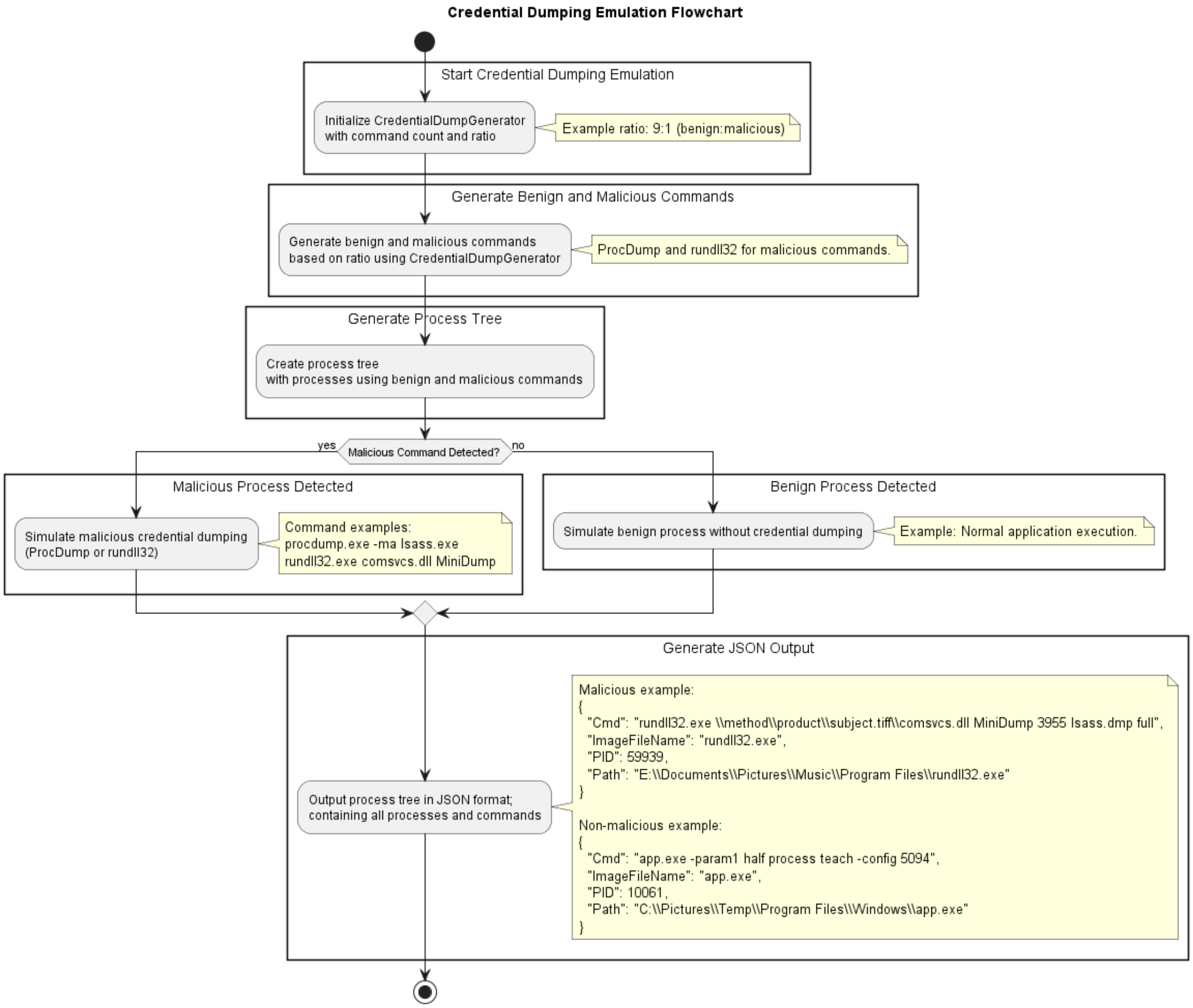

**Fig. 11:** Credential Dumping Emulation Flowchart

*8) Results Benchmarking:* The benchmarking results for SPECTRE's modules demonstrate efficient performance, scalability, and resource utilization. SPECTRE scales efficiently, with linear processing times in emulation and moderate memory usage in delta analysis and combined plotting, making it suitable for large forensic investigations. The system exhibits effective memory management with modest growth, particularly in emulation processes, which optimize performance in data-heavy scenarios. SPECTRE ensures consistent performance in real-time forensics, with minimal increases in processing time for time-critical operations like timeline and memory analysis plotting.

In terms of trade-offs, there are some challenges related to resource demand and scalability, particularly with anomaly detection and emulation when working with large datasets, which may be a limitation in resource-constrained environments. Additionally, the high resource requirements for creating realistic training environments can impact the system's ability to efficiently handle real-time incident response.

Overall, SPECTRE effectively balances scalability, efficiency, and resource utilization. However, there are opportunities to optimize resource-heavy modules, such as anomaly detection and delta analysis, to further enhance performance in resource-limited settings. Table IV provide a comparative analysis of SPECTRE with current industry and related research systems.

### E. SPECTRE Overall Evaluation

The evaluation of SPECTRE across various metrics confirms its robustness:

1) **Throughput:** SPECTRE demonstrates high throughput, with linear scaling in emulation, ensuring it handles large datasets efficiently.
2) **Latency:** Minimal latency in critical modules ensures SPECTRE is suitable for real-time incident response.
3) **Accuracy:** SPECTRE achieves 100% accuracy in anomaly detection, validating the effectiveness of its algorithms.

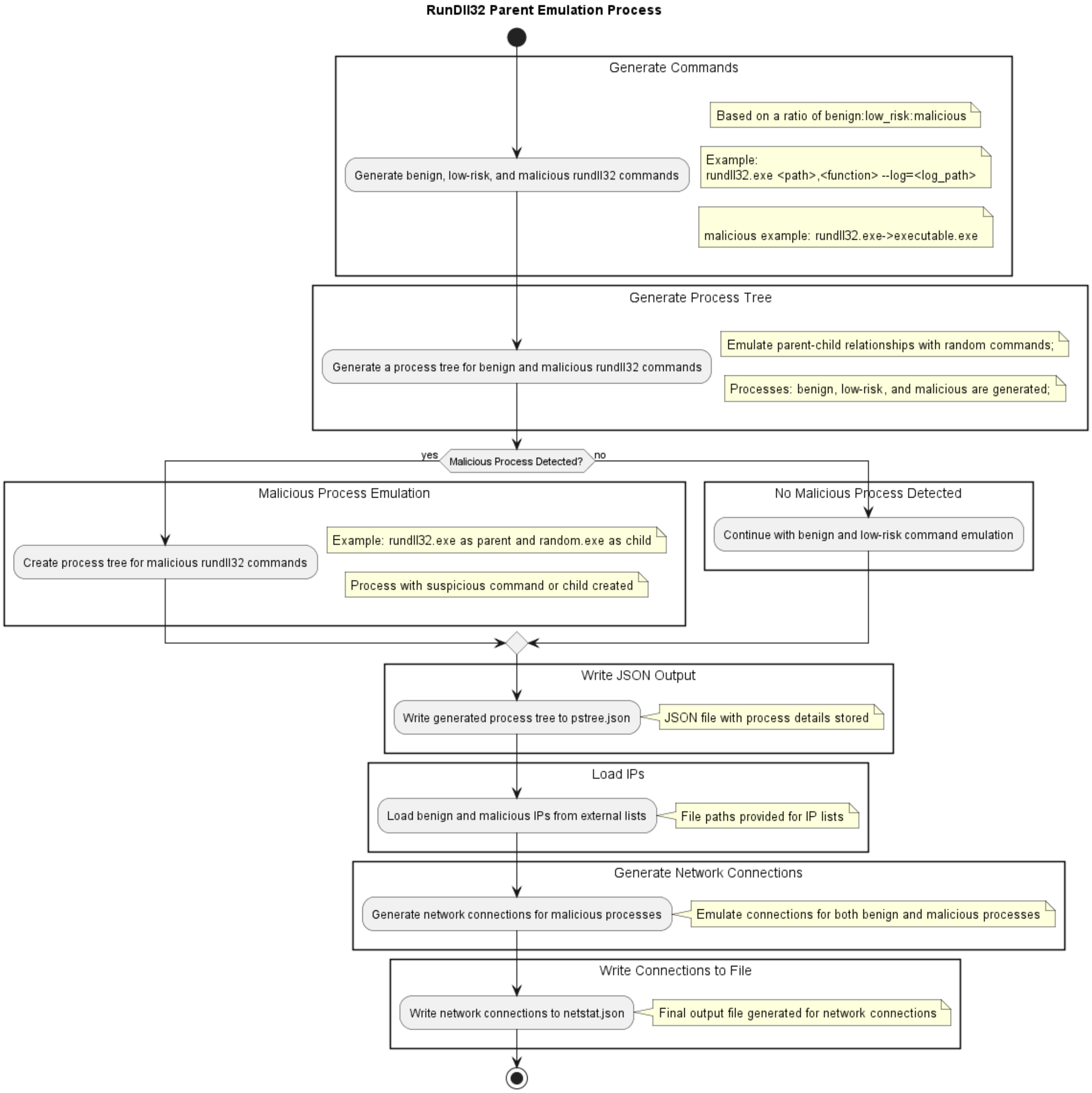

**Fig. 12:** RunDLL32.exe Malicious Process Emulation Flowchart

4) **Scalability:** The system shows near-linear scalability in processing time and memory usage as datasets grow, demonstrating its capability to manage increasing forensic data.
5) **Resource Utilization:** SPECTRE maintains efficient resource usage across most modules, with the highest memory usage observed in anomaly analysis, which remains manageable even with larger datasets.

## V. Results Validation and Discussion

### A. Strengths of SPECTRE

SPECTRE represents a groundbreaking advancement in memory forensics, addressing key challenges in cybersecurity with innovative capabilities. By equipping analysts with cutting-edge framework, it enhances threat detection, response, and collaboration. This section explores SPECTRE's transformative features and contributions in detail.

*1) Visualizations:* SPECTRE's advanced visualizations empower cybersecurity analysts by offering comprehensive insights into system behavior through memory analysis plots, delta analysis charts, and timeline visualizations. These tools facilitate the detection of anomalies such as excessive child processes, unusual connection patterns, abrupt changes in processes or connections, and zero-day vulnerabilities, effectively uncovering threats like process injection and data exfiltration. Additionally, temporal correlation of processes and connections aids in tracking security events, identifying root causes, and enhancing response efficiency, making the platform a critical asset for proactive threat mitigation.

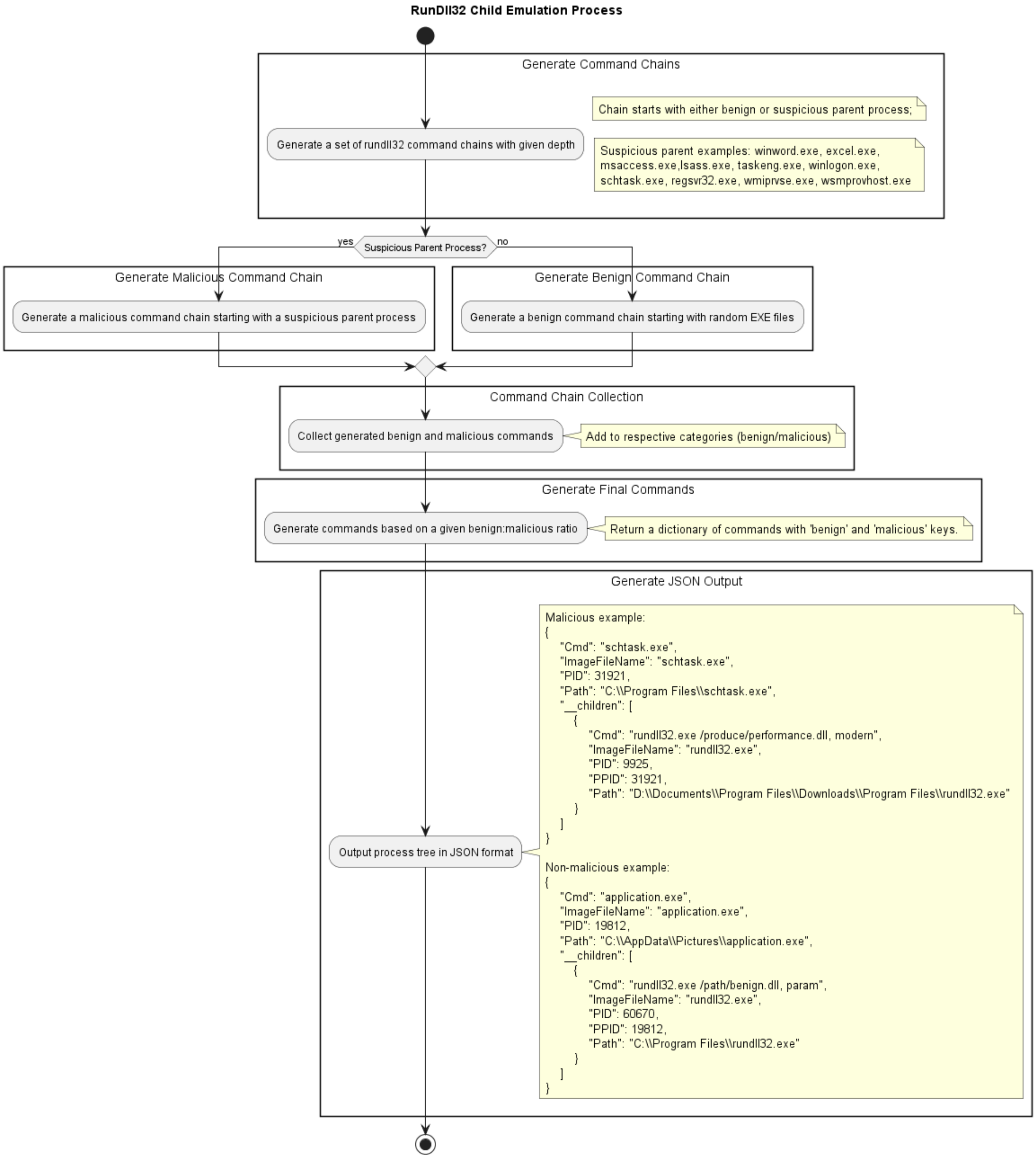

**Fig. 13:** RunDLL32.exe Malicious Child Emulation Flowchart

*2) Intermediate Format:* The adoption of Volatility's JSON format allows SPECTRE to achieve seamless integration with other forensic systems. By using a standardized output, it streamlines data processing, minimizes manual intervention, and ensures compatibility with both emulated and real-world memory dumps. This enhances operational efficiency and supports consistent workflows in memory forensics.

*3) Test Data Emulation:* SPECTRE's emulation capabilities allow the realistic simulation of advanced attack scenarios, such as credential dumping and malicious connection. These simulations provide controlled environments for system testing, skill development, and response planning, while also validating detection mechanisms and refining forensic workflows to enhance overall security preparedness.

*4) Anomaly Detection:* The anomaly detection module targets critical areas, including credential dumping, malicious IP detection, and suspicious process behavior. It identifies advanced malware tactics like RunDLL32 exploitation, offers actionable insights for combating stealthy attacks, and enhances organizational readiness through testing in controlled environments.

**TABLE IV:** Comparison of SPECTRE, MemProcFS [60], [61], and Volatility Workbench

| Feature | MemProcFS | SPECTRE | Volatility Workbench |
|---|---|---|---|
| **Visualization Aspect** | Represents raw memory as files within a virtual file system. | Provides detailed visualizations of raw memory to facilitate analysis. | Provides a graphical interface for executing Volatility commands [34]. |
| **Malicious Behavior Detection** | • Identifies malicious behaviours using the `findevil` built-in plugin.<br>• Triggers `PROC_PARENT` if a known process is associated with a suspicious parent process.<br>• Flags `PE_HDR_SPOOF` for modules employing spoofing techniques in their PE headers.<br>• Does not provide IP-related detections. | • Identifies malicious behaviours through the Anomaly Detection Module.<br>• Detects malicious `RunDLL32` child processes when associated parent processes are invalid.<br>• Identifies spoofing attempts using the Malicious Extension Detector.<br>• Supports detection of compromised and safe IPs, with integrated VirusTotal lookup, geolocation, and WHOIS services. | Volatility output has to be manually observed to find any anomalies, no additional malicious behavior detection is supported by Volatility Workbench |
| **Emulation Support** | Not supported. | Supports emulations of processes, network connections, users, registries, and modules via Volatility JSON. | Not supported |
| **Delta Analysis** | Not supported. | Offers delta analysis for processes, connections, registries, users, and modules across memory dumps. | Does not support delta analysis. |
| **Timeline Functionality** | • Displays all common forensic elements in chronological order for a single memory dump.<br>• Does not support the analysis of multiple memory dumps.<br>• Does not associate processes with their corresponding network connections in the timeline, requiring manual analysis. | • Provides a timeline for processes and connections within a single memory dump.<br>• Supports timeline of multiple memory dumps and generates comprehensive timelines for processes, users, modules, registries, and connections.<br>• Associates connection counts with processes in the timeline and links malicious IPs via scatter plot visualizations. | Manual interpretation of raw Volatility data is required to perform the analysis |
| **Multi-Language Support** | Supports APIs for C/C++, C#, Java, Rust, and Python. | Provides support for Python. | Primarily a GUI software, no language API is supported. |

*5) IP Forensics:* The IP forensics module enhances memory analysis with external intelligence through Whois lookups, geolocation, blacklist checks, and VirusTotal queries. It correlates memory anomalies with network threats, enabling strategic security responses based on IP-level intelligence.

*6) Red, Blue, and Purple Team Applications:* SPECTRE empowers cybersecurity teams with tailored functionalities: Red Teams leverage its capabilities to simulate attacks and test vulnerabilities, Blue Teams utilize it to detect and mitigate threats effectively, and Purple Teams harness its features to enhance collaboration through shared visualizations and emulated scenarios.

### *B. Comparison with Existing Literature*

SPECTRE extends and enhances prior research in memory forensics and malware analysis by building upon the Volatility framework with added features like snapshot comparison, emulation, and advanced visualization. It addresses critical gaps in behavior-based malware detection by concentrating on memory manipulations and integrating network-contextualized intelligence with traditional forensic approaches, significantly improving threat investigation processes.

### *C. Novel Contributions*

SPECTRE introduces groundbreaking innovations, including transforming raw forensic data into actionable insights through intuitive visualizations, enabling real-time attack scenario simulation for training and stress testing via emulation, and ensuring interoperability across platforms like VolMemLyzer and volGPT through its JSON-based design. These advancements underscore its value in modern cybersecurity practices.

### *D. Limitations and Future Work*

While SPECTRE has demonstrated its utility in detecting and analysing malware through synthetic data and anomaly detection, several enhancements can further expand its capabilities. future enhancements include:

- **Document Analysis:** Integrating document analysis features to detect document-based malware.
- **Snapshot Tagging:** Incorporating tags for snapshot generation to streamline emulated environment creation.
- **Non-JSON Outputs:** Supporting non-JSON output formats from Volatility and incorporating additional commands to enhance compatibility and extend analysis capabilities.
- **Efficiency in Process Trees:** Converting `pstree` output to a dictionary format using PID as keys to achieve $O(n)$ efficiency and supporting user-specified root process reassignment.

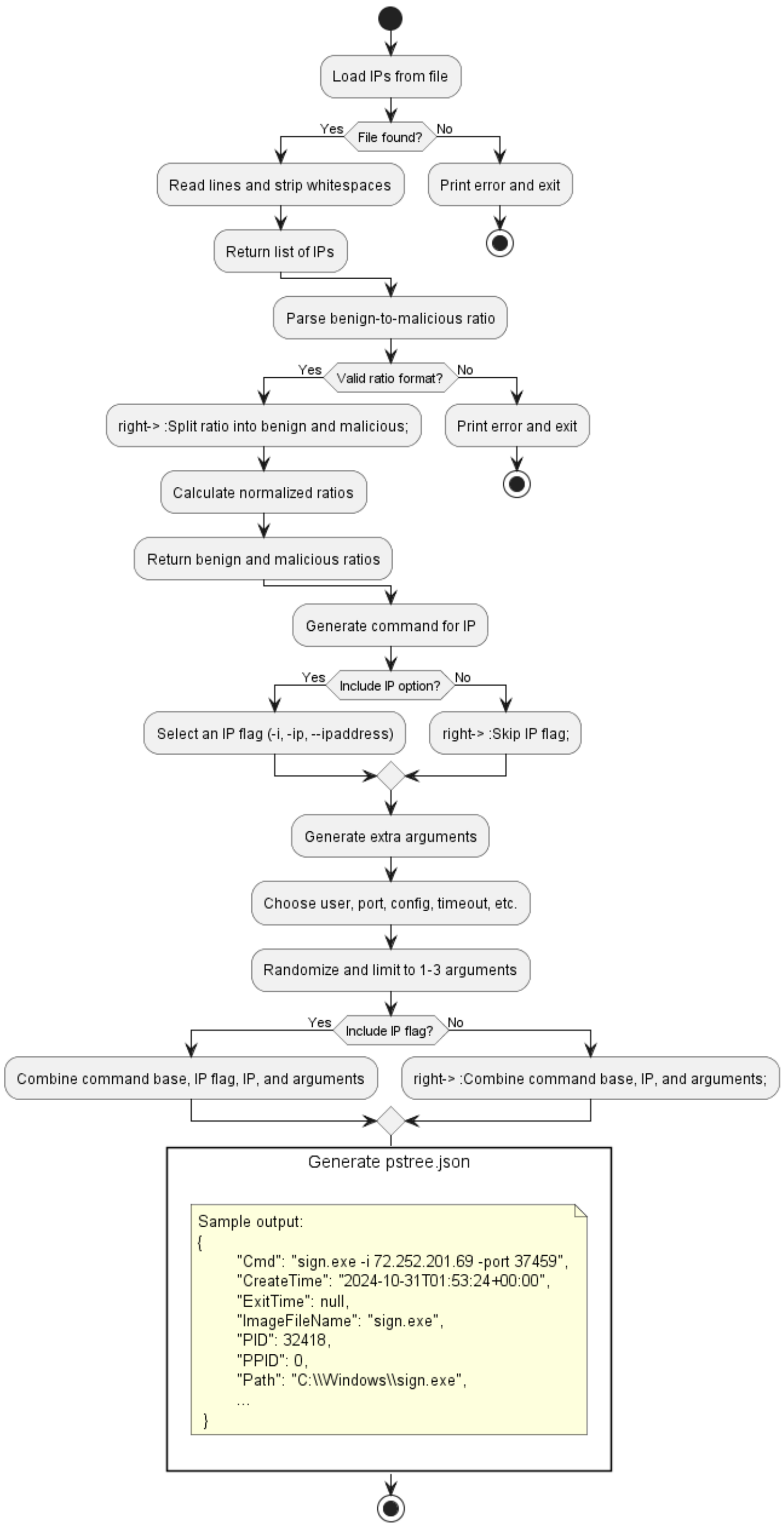

**Fig. 14:** Command-line IP Emulation Flowchart

- **DLL Injection Detection:** Enhancing detection using additional keys from the `ldrmodules` plugin.
- **Pattern Recognition:** Incorporating regular expressions to identify patterns for anomaly detection and malware emulation.
- **HTML Output:** Expanding output formats to include HTML-based visualizations for better accessibility and reporting.
- **File Dumping:** Utilizing file dumping features to detect malware via executable hashes.
- **Offset Retrieval:** Developing methods to use offsets for data retrieval in automated memory dumps.
- **Cross-Platform and Multi OS Memory Dump Support:** Expanding compatibility to include Linux and other operating systems.

## VI. Conclusion

This paper presented a novel approach to digital forensic analysis through the development of a memory data emulation system, SPECTRE. By leveraging emulated data and synthetic process generation, SPECTRE facilitates controlled, efficient, and reproducible testing for forensic analysis methods, addressing a critical need in modern cybersecurity research. Key contributions of this work include the design and implementation of modules for generating realistic scenarios, which mimic the behaviours of benign and malicious activities. These modules provide flexibility in defining custom scenarios, enabling robust testing of forensic capabilities without the need for live malware samples. Additionally, SPECTRE supports compatibility with industry-standard systems like Volatility by producing JSON-formatted outputs for seamless integration into established workflows. The use of synthetic data creation highlights the versatility of the system. This combination not only provides a diverse dataset for testing but also avoids the risks associated with handling actual malware. Moreover, the system aligns with the broader goals of memory analysis, delta analysis, and timeline reconstruction. To sum upn, SPECTRE provides a valuable contribution to the field of digital forensics by enabling safer, more flexible, and more accessible testing of forensic analysis systems and methodologies. It lays a strong foundation for advancing forensic research and equips investigators with systems to adapt to evolving cyber threats in a controlled and ethical manner.

## Footnotes

### *Ethical Approval*

This research was determined to not require ethical approval.

### *Competing Interests*

The authors confirm that they have no competing interests or personal relationships that could have influenced the work presented in this paper.

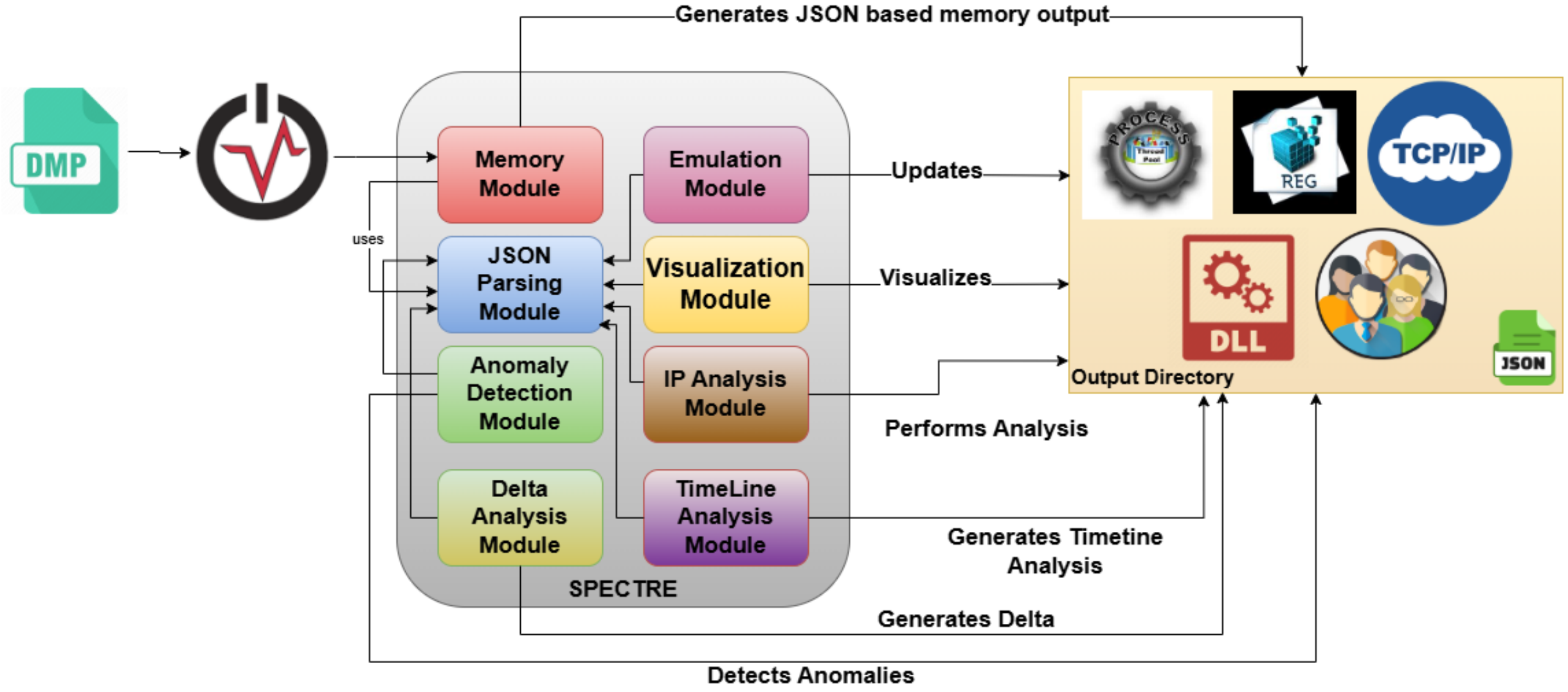

**Fig. 15:** SPECTRE Detection, Response and Investigation Diagram

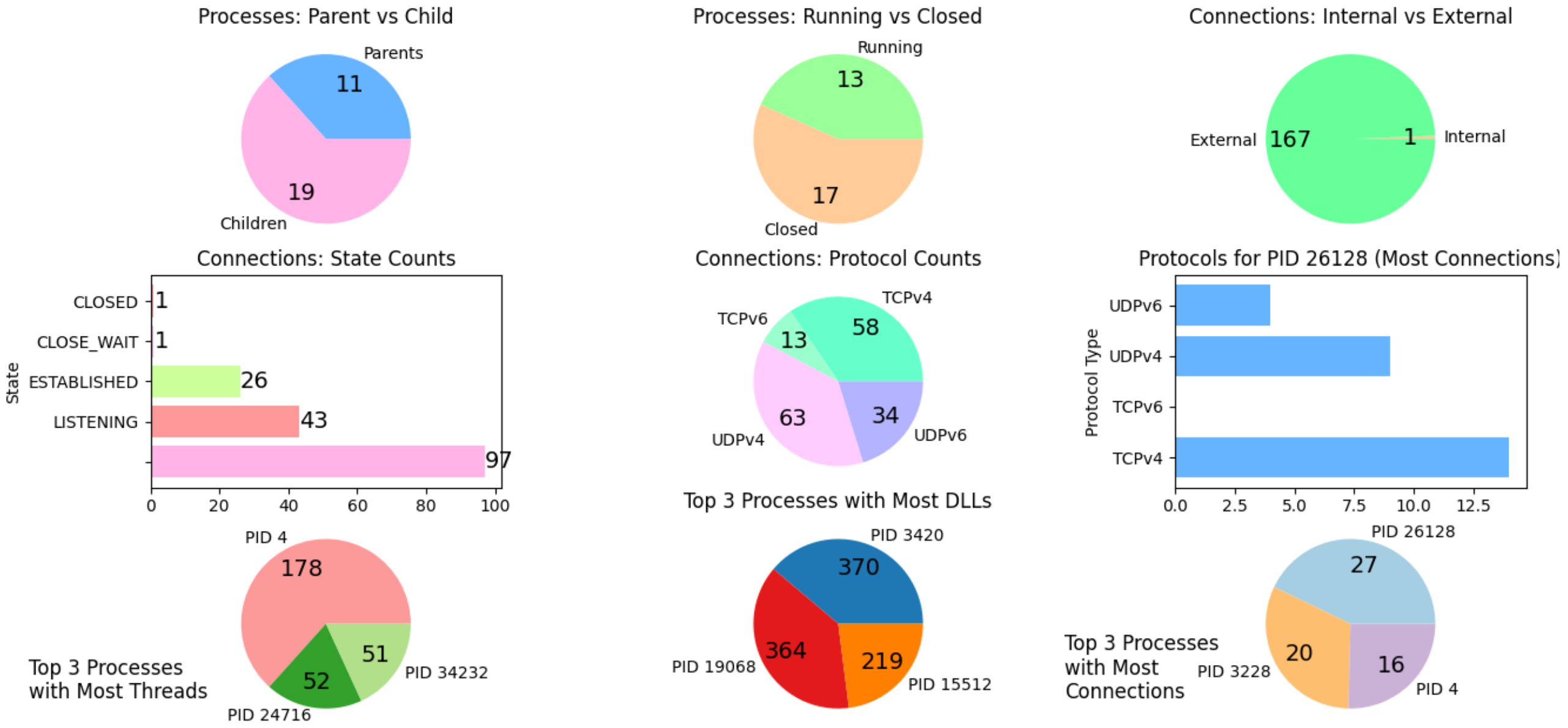

**Fig. 16:** SPECTRE Memory Analysis Plot

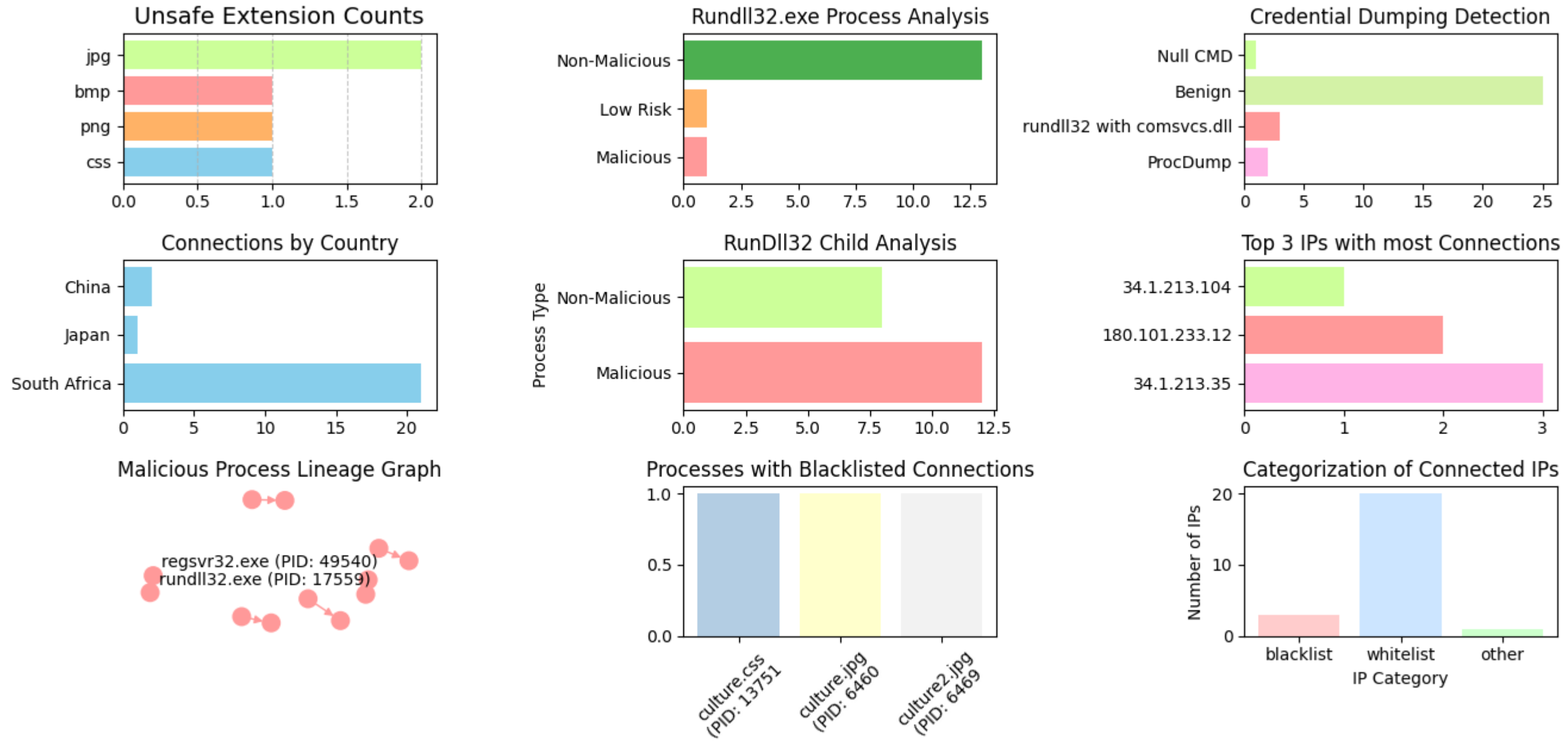

**Fig. 17:** SPECTRE Anomaly Analysis Plot

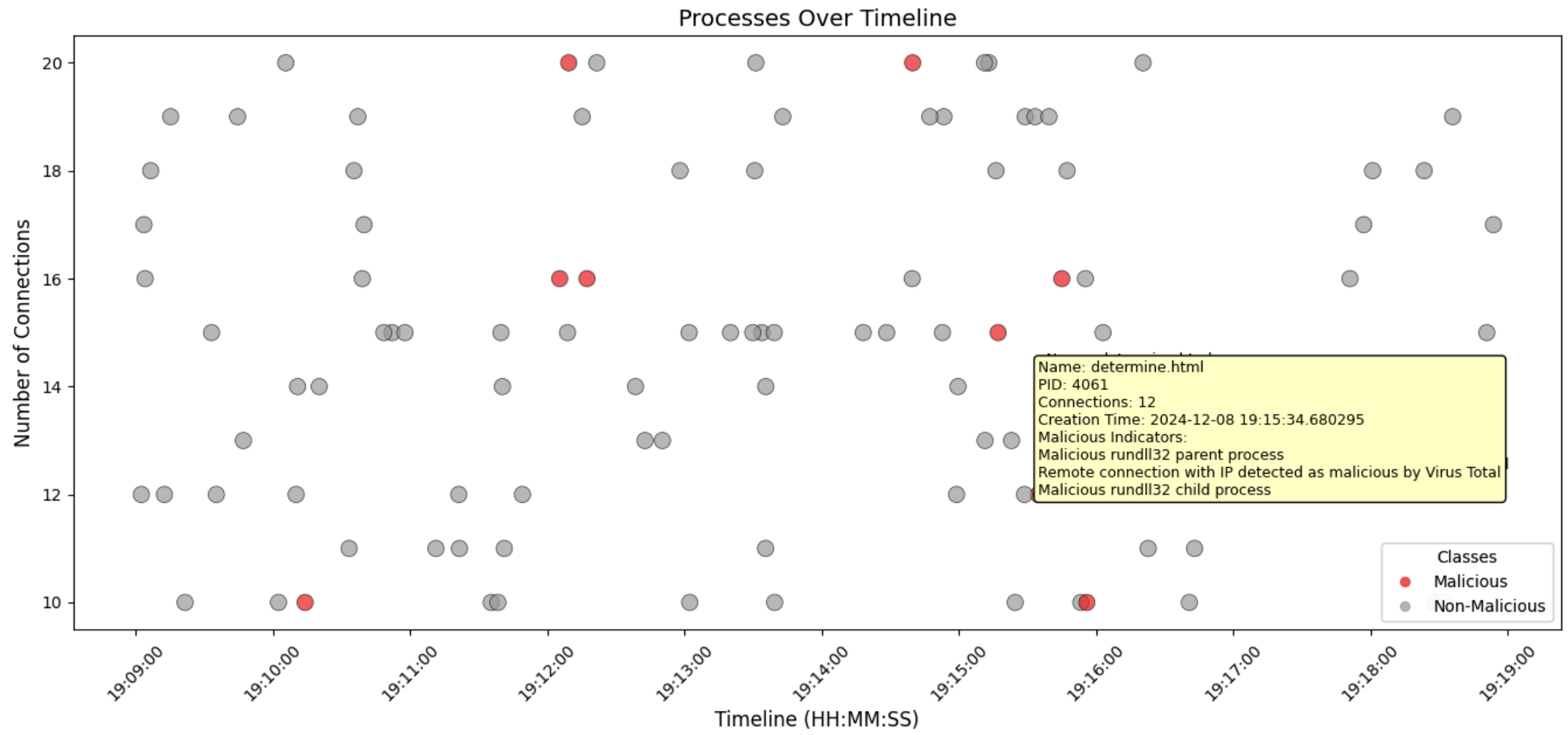

**Fig. 18:** Detailed Processes Scatter Plot

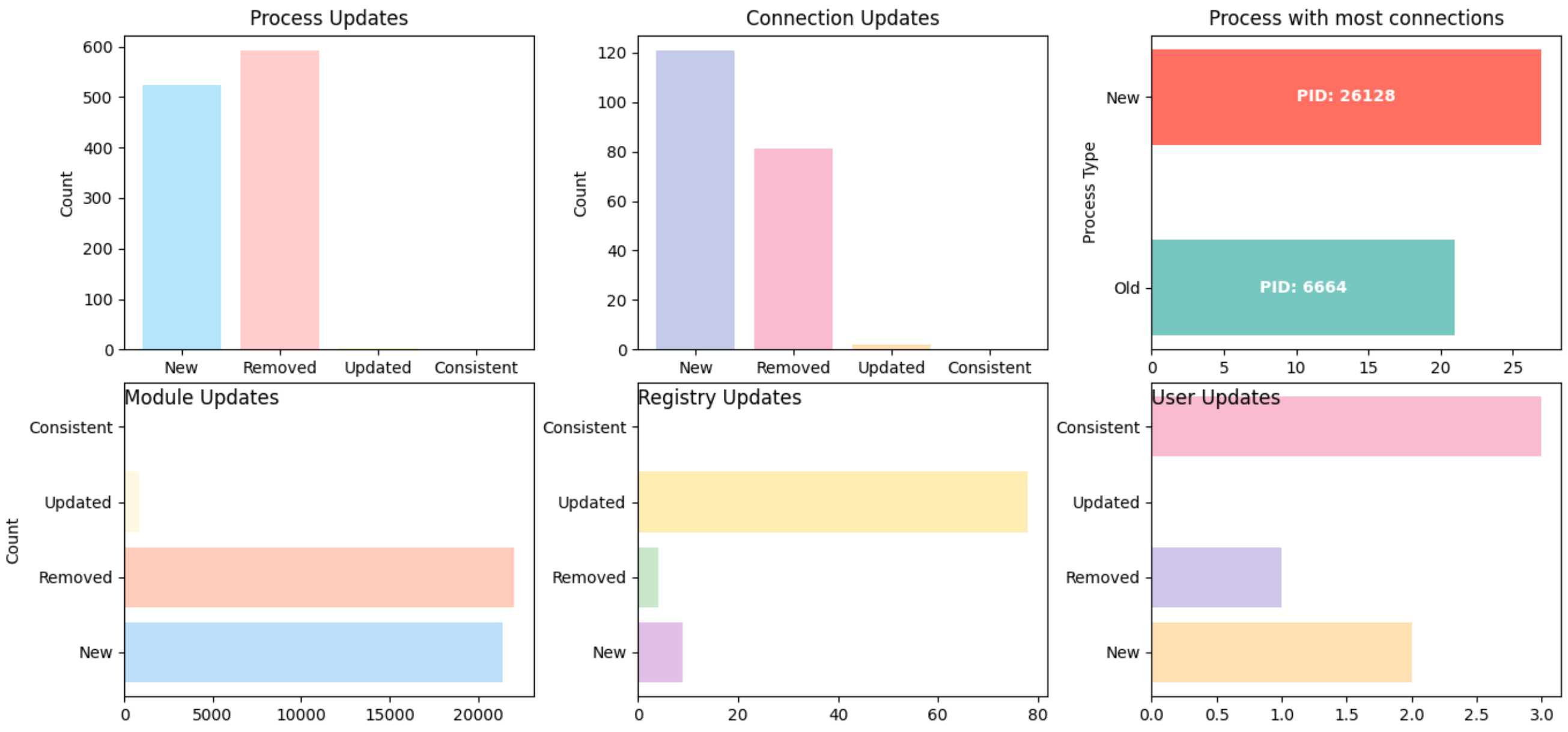

**Fig. 19:** Delta Analysis Plot

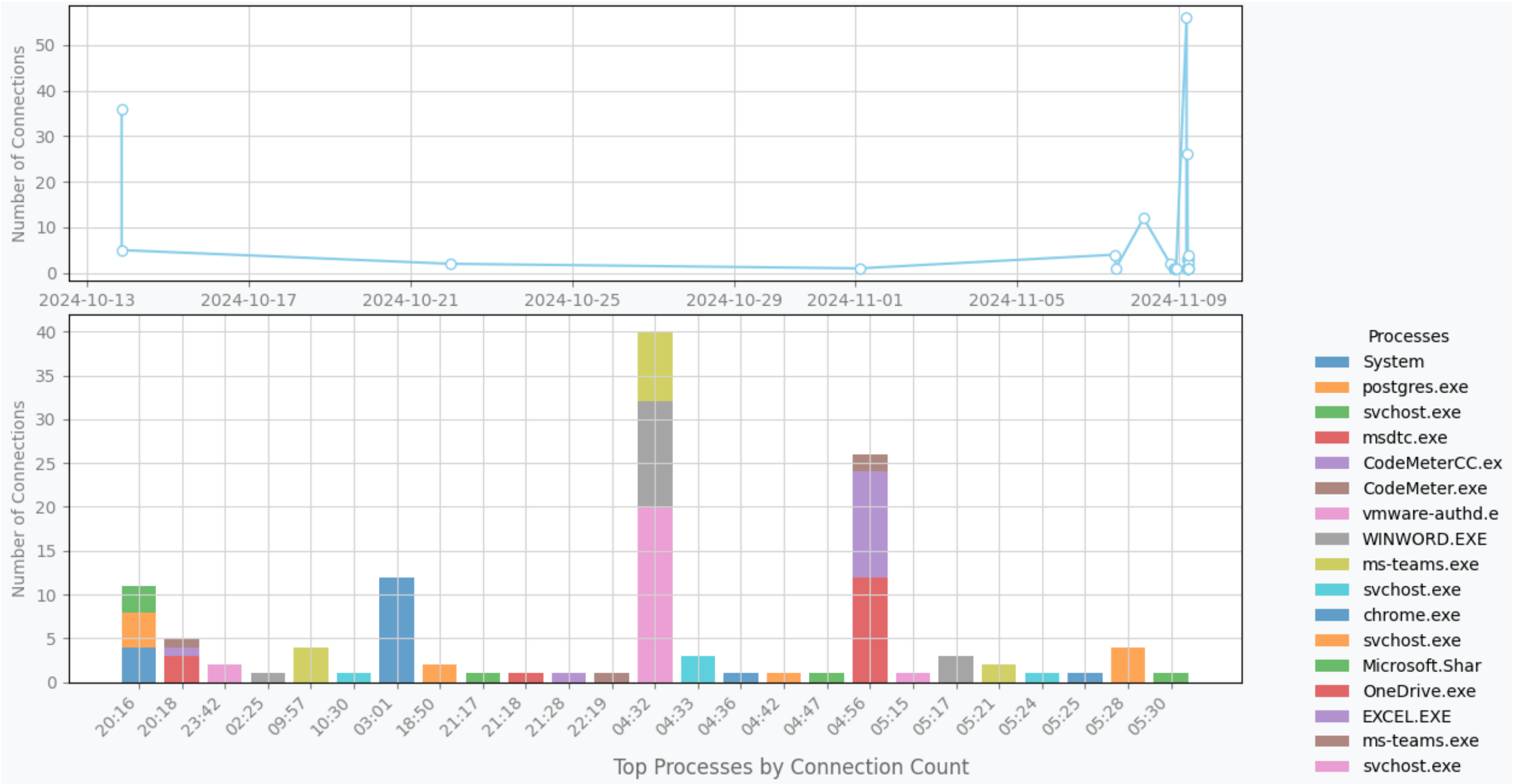

**Fig. 20:** Processes and Connections Timeline Plot

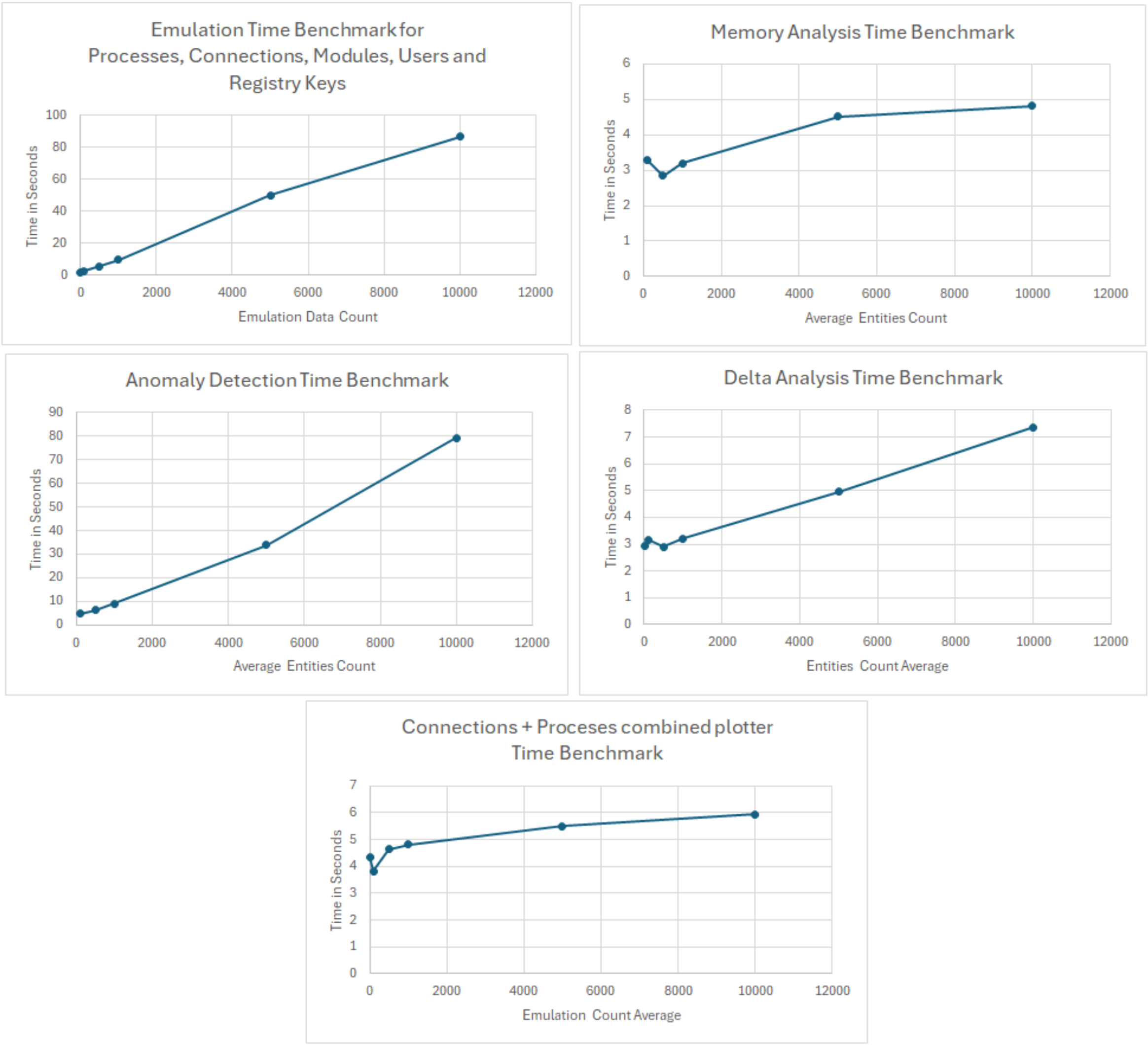

**Fig. 21:** Time Benchmarking Results

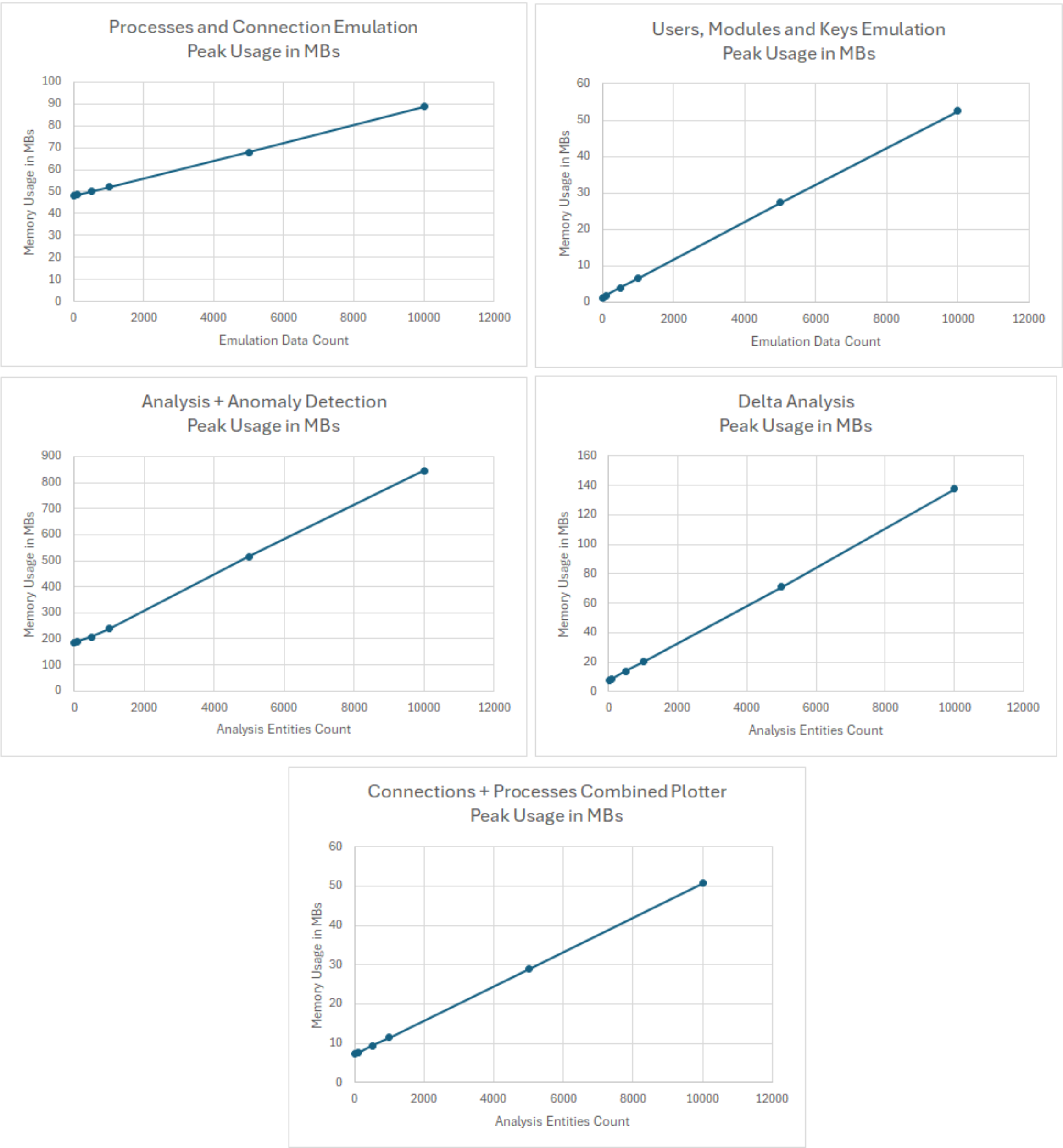

**Fig. 22:** Memory Benchmarking Results